\documentclass[trackchanges, onecolumn]{aastex7}
\usepackage{amsmath}
\usepackage{graphicx} 
\usepackage{tikz}
\usetikzlibrary{arrows.meta,positioning,fit,calc}

\begin{document}

\title{Reconstructing the Stripping History of the Sagittarius Stream with Neural Networks}
 
\author{Jian Zhang}
\affiliation{School of Astronomy and Space Sciences,
University of Chinese Academy of Sciences,
Beijing 100049, People's Republic of China}
\email{zhangjian191@mails.ucas.ac.cn}

\author{Cuihua Du}
\affiliation{School of Astronomy and Space Sciences,
University of Chinese Academy of Sciences,
Beijing 100049, People's Republic of China}
\email[show]{ducuihua@ucas.ac.cn}

\author{Mingji Deng}
\affiliation{School of Astronomy and Space Sciences,
University of Chinese Academy of Sciences,
Beijing 100049, People's Republic of China}
\email{dengmingji22@mails.ucas.ac.cn}

\author{Zhongcheng Li}
\affiliation{School of Astronomy and Space Sciences,
University of Chinese Academy of Sciences,
Beijing 100049, People's Republic of China}
\email{lizhongcheng20@mails.ucas.ac.cn}

\author{Haoyang Liu}
\affiliation{School of Astronomy and Space Sciences,
University of Chinese Academy of Sciences,
Beijing 100049, People's Republic of China}
\email{liuhaoyang23@mails.ucas.ac.cn}

\begin{abstract}
The Sagittarius (Sgr) Stream is produced by the ongoing disruption of the Sgr dwarf spheroidal (dSph) galaxy and is thought to contain multiple wraps that were stripped during different pericentric passages. In this study, we introduce a neural‐network–based method trained on N-body simulations to infer the stripping time of Sgr Stream stars directly from their phase-space coordinates. We combine spectroscopic data from SEGUE, APOGEE DR17, and LAMOST DR7 LRS with \textit{Gaia} EDR3 astrometry and distance estimates from the latest \texttt{StarHorse} catalog to identify high-quality Sgr Stream members. Applying our method to these stars, we measure a clear metallicity gradient with stripping time, well described by a linear relation with slope $\sim 0.3~\mathrm{dex~Gyr^{-1}}$. We further predict the stripping times of globular clusters previously suggested to originate from the Sgr dSph. M 54, Terzan 7, Terzan 8, and Arp 2 exhibit stripping times consistent with being currently bound to the Sgr remnant. Pal 12, Whiting 1, and NGC 2419 are inferred to have been stripped $0.9 \pm 0.1$, $1.1 \pm 0.2$, and $2.1 \pm 0.2$ Gyr ago, respectively. For NGC 4147 and NGC 5634, whose membership in the Sgr system remains uncertain, our analysis suggests stripping times of $1.1 \pm 0.4$ and $1.1 \pm 0.1$ Gyr, respectively, if they are ultimately confirmed as genuine Sgr members. These results demonstrate that data-driven models of dynamical stripping histories offer a promising approach for reconstructing the formation and chemical evolution of the Sgr Stream.

\end{abstract}


\section{Introduction} 
The $\Lambda$CDM paradigm predicts that large-scale structures grow hierarchically through the continual accretion and merging of smaller systems \citep{White78, Springel05}. This is clearly reflected in the Milky Way, whose stellar halo contains numerous substructures, such as the Helmi Streams \citep{Helmi99}, Gaia–Sausage–Enceladus \citep{Belokurov18, Helmi18}, the Sgr Stream \citep{Ibata94, Majewski03}, Thamnos \citep{Koppelman19}, Sequoia \citep{Myeong19}, and Arjuna, I'itoi, and Aleph \citep{Naidu20}. Among these, the Sgr Stream is particularly noteworthy because it represents an ongoing merger: the tidal disruption of the Sgr dSph, the Milky Way’s nearest and third most massive satellite. Studying the Sgr Stream, and its interaction with the Milky Way, provides a unique opportunity to investigate the detailed physics of how galactic mergers unfold.

Stellar streams represent the remnants of stars stripped from their progenitors by the tidal field of the Milky Way. They manifest as narrow, coherent structures on the sky. Streams born from globular clusters are typically thin and dynamically cold, whereas those arising from dwarf galaxies are broader and dynamically hotter \citep{Bonaca25}. The Sgr Stream is the archetype of the latter type, produced by the ongoing tidal disruption of the Sgr dSph.

The Sgr dSph was first recognized by \citet{Ibata94} as an overdensity in both position and velocity, and subsequent studies further mapped its outskirts and extended structure \citep{Alard96, Mateo96, Mateo98}. Since then, numerous studies have used a variety of stellar tracers---including RR Lyrae stars, blue horizontal branch stars, M giants, and main-sequence turnoff stars---together with techniques such as matched color--magnitude diagram (CMD) filtering and great-circle cell counts (GC3) to trace the tidal debris of the Sgr Stream across the sky \citep[e.g.,][]{Ibata01, Majewski03, Newberg03, Belokurov06, Niederste10, Correnti10, Boer14, Belokurov14, Hernitschek17}. In parallel, dynamical modeling efforts, ranging from early N-body simulations \citep{Helmi01, Helmi04, Johnston05, Law05, Fellhauer06, Penarrubia10} to the landmark study of \citet{Law10}, have progressively improved our understanding of the formation and evolution of the Sgr Stream.

The \textit{Gaia} mission has provided precise astrometric measurements for more than a billion stars in the Milky Way \citep{Gaia16a, Gaia16b, Gaia18, Gaia21}, fundamentally transforming our understanding of Galactic structure. For the Sgr Stream in particular, \textit{Gaia} has enabled the blind identification of $\sim 10^5$ high-probability members based primarily on proper motions \citep{Antoja20, Ibata20, Ramos22}. These data have significantly improved our knowledge of the Sgr Stream's present-day kinematics and enabled the construction of comprehensive dynamical models of the Sgr dSph and the Sgr Stream \citep{Vasiliev21, Oria22, Wang22}.

The chemical properties of stars in the Sgr dSph and the Sgr Stream provide important clues to the nature of the progenitor dwarf galaxy. Numerous studies have shown that stars in the Sgr Stream ([Fe/H] $\sim -1$) are, on average, more metal-poor than those in the bound core of the Sgr dSph ([Fe/H] $\sim -0.4$) \citep{Bellazzini06, Chou07, Monaco07, Carlin12}. Metallicity gradients have been detected both within the Sgr dSph core \citep{Majewski13, Mucciarelli17, Vitali22} and along the Sgr Stream arms \citep{Boer14, Boer15, Cunningham24, Muraveva25}, suggesting that the Sgr progenitor possessed a radial metallicity gradient prior to its tidal disruption.

However, measuring a metallicity gradient along the Sgr Stream is complicated by the overlap of multiple tidal wraps. Older wraps, stripped early from the progenitor's outer regions, are expected to be more metal-poor, whereas younger wraps, removed from deeper layers, should be more metal-rich. The superposition of these wraps obscures any simple metallicity–distance or metallicity–longitude trend. 

The multi-wrap nature of the Sgr Stream has been substantiated in several recent studies. \citet{Gibbons17} used SEGUE \citep{Yanny09a}  spectroscopy to identify two chemically and kinematically distinct sub-populations in both the leading and trailing arms. \citet{Johnson20}, using the H3 Survey \citep{Conroy19}, found a kinematically diffuse, metal-poor population that they attributed to the stellar halo of the Sgr progenitor. Most recently, \citet{Limberg23} combined SEGUE spectroscopy with \textit{Gaia} EDR3 astrometry to show that simulated particles stripped more than 2 Gyr ago exhibit present-day phase-space properties similar to low-metallicity Sgr Stream stars, whereas those stripped within the last 2 Gyr more closely resemble metal-rich ([Fe/H] $>$ –1) Sgr Stream members.

In this work, we address the challenge posed by overlapping tidal wraps by introducing a neural-network method, trained on N-body simulations, to infer the stripping time of Sgr Stream stars directly from their phase-space coordinates. This paper is organized as follows. Section~\ref{sec:data} describes the observational datasets and the selection criteria used to identify Sgr Stream members. Section~\ref{sec:methods} introduces our neural-network framework for mapping phase-space coordinates to stripping time. Our results on the metallicity gradient are presented in Section~\ref{sec:results}. Section~\ref{sec:discussion} discusses the implications and limitations of our findings, and Section~\ref{sec:conclusion} summarizes our main conclusions.

\section{Data}\label{sec:data}
\subsection{SEGUE/APOGEE/LAMOST + \textit{Gaia} and \texttt{StarHorse}}
This study extends the analysis presented in \citet{Limberg23} (hereafter L23). L23 combined spectroscopic data from SEGUE, which is well suited for studies of distant halo substructures \citep{Yanny09b, Belokurov14, Gibbons17, Yang19}, with astrometry from \textit{Gaia} EDR3 \citep{Gaia21}. Although \textit{Gaia} provides exceptionally precise astrometric measurements, its parallaxes offer only limited constraints for stars in the Sgr Stream due to its large heliocentric distance \citep{Ramos22}. To mitigate this limitation, L23 adopted distance estimates from the latest \texttt{StarHorse} catalog \citep{Queiroz23}, which was generated using the Bayesian isochrone-fitting code \texttt{StarHorse}.

The latest \texttt{StarHorse} catalog provides homogeneous distance and extinction estimates for more than ten million stars observed by several major spectroscopic surveys, including GALAH+ DR3 \citep{DeSilva15, Martell17}, LAMOST DR7 \citep{Cui12, Zhao12}, APOGEE DR17 \citep{Abdurro'uf22}, RAVE DR6 \citep{Steinmetz20}, SDSS DR12 (including SEGUE and BOSS) \citep{Yanny09a, Alam15}, Gaia--ESO DR5 \citep{Gilmore12, Randich22}, and the \textit{Gaia} RVS sample from \textit{Gaia} DR3 \citep{Gaia23}. After applying the Sgr Stream selection criteria described in Section~\ref{subsec:select} following \citet{Limberg23}, we find that only SEGUE, APOGEE DR17, and LAMOST DR7 LRS contain a sufficiently large number of Sgr Stream stars for a statistically robust analysis. We therefore restrict our study to these three surveys. For a comprehensive description of the \texttt{StarHorse} methodology and the full set of spectroscopic inputs, we refer the reader to \citet{Queiroz23} (hereafter Q23).

Q23 apply survey-specific selection criteria tailored to the characteristics of each spectroscopic survey. SEGUE predominantly contains stars with $T_{\rm eff}$ in the range 4000--10000\,K and spectral signal-to-noise ratios greater than 10 \citep{Lee08}. For this survey, Q23 adopt the recommended stellar atmospheric parameters ($T_{\rm eff}$, $\log g$, and [Fe/H]) from SDSS DR12. They impose a stricter signal-to-noise ratio threshold of 20 and require all three atmospheric parameters to be available. 
For APOGEE DR17, Q23 adopt the $T_{\rm eff}$, $\log g$, and [Fe/H] from the ASPCAP pipeline \citep{GarciaPerez16, Jonsson20}, retaining only stars with available $H_{2MASS}$ passband and reliable spectroscopic parameters. 
For LAMOST DR7, Q23 adopt stellar atmospheric parameters ($T_{\rm eff}$, $\log g$, and [Fe/H]) from the LASP pipeline \citep{Wu14}. They require available 2MASS $K_s$ photometry and apply the following uncertainty cuts: $\sigma_{T_{\rm eff}}<500\,\mathrm{K}$, $\sigma_{\log g}<0.8\,\mathrm{dex}$, and $\sigma_{\mathrm{[Fe/H]}}<0.5\,\mathrm{dex}$. 

We cross-matched each of the three surveys with \textit{Gaia} EDR3 with a search radius of $1''$. We then incorporated \texttt{StarHorse} distances by matching the corresponding \textit{Gaia} EDR3 source IDs. To ensure high-quality astrometry, we required that the \textit{Gaia} EDR3 Renormalized Unit Weight Error (RUWE) $\leq 1.4$ \citep{Lindegren21} and the parallax be greater than $-5$\,mas \citep{Queiroz23}. Furthermore, we limited our sample to metal-poor stars ($\mathrm{[M/H]} < -0.5$, where $\mathrm{[M/H]}$ is taken from \texttt{StarHorse}; unless otherwise noted, all metallicities used below refer to the \texttt{StarHorse} values), a selection that effectively minimizes contamination from the thin disk while retaining the majority of Sgr Stream members \citep{Hayes20, Limberg23}. Finally, to ensure reliable distance estimates, we selected sources with relative distance uncertainties of less than 20\%.

\subsection{Selection of Sgr Stream Members}\label{subsec:select}
We use positions (\(\alpha, \delta\)) and proper motions (\(\mu^*_\alpha, \mu_\delta\)) from \textit{Gaia} EDR3. Line-of-sight velocities \(v_{los}\) are from SEGUE, APOGEE DR17, and LAMOST DR7 LRS, and heliocentric distances ($d_\odot$) are adopted from the \texttt{StarHorse} catalog. These six-dimensional phase-space coordinates are transformed into the Galactocentric Cartesian coordinates using the \texttt{Astropy} \citep{Astropy22} package, assuming a Solar position of $(x_\odot, y_\odot, z_\odot) = (8.122,\;0,\;0.0208)\,\mathrm{kpc}$ \citep{GRAVITY18, Bennett19} and a Solar velocity of $(U_\odot, V_\odot, W_\odot) = (11.1,\;245,\;7.25)\,\mathrm{km\,s^{-1}}$ \citep{Schonrich10, McMillan17}. 

We compute the angular momentum components (\textbf{L}) directly from the Galactocentric Cartesian coordinates. Actions \(\mathbf{J} = (J_R, J_\phi, J_z)\) are calculated using the \texttt{Agama} package \citep{Vasiliev19} within the axisymmetric Galactic potential of \citet{McMillan17}. To propagate observational uncertainties, we generate 100 Monte Carlo (MC) realizations for each star by sampling its proper motions, line-of-sight velocity, and distance within their respective uncertainties. We adopt the median values of the resulting distributions as the final orbital parameters, with the 16th and 84th percentiles serving as uncertainty estimates.

Following \citet{Limberg23}, we apply a set of selection criteria designed to obtain a clean sample of Sgr Stream candidates:

(1) \(J_z > J_R\);

(2) \(d_\odot > 6~\mathrm{kpc}\);

(3) \(-10 < L_y/(10^3~\mathrm{kpc~km~s^{-1}}) < -3\);

(4) \(-4 < L_z/(10^3~\mathrm{kpc~km~s^{-1}}) < +1\).

After applying these criteria, we identify $\sim 1600$ Sgr Stream candidates in SEGUE, $\sim 1400$ in LAMOST DR7 LRS, and $\sim 1700$ in APOGEE DR17. These samples form the basis of our analysis of the Sgr Stream's metallicity gradient.

\section{Methods}\label{sec:methods}
In this section, we describe the neural-network (NN) framework used to predict the stripping time of Sgr Stream stars from their phase-space coordinates. A NN consists of a sequence of interconnected layers, each of which applies an affine transformation followed by a nonlinear activation. For a given layer, the transformation can be written as
\(Z = \Phi\left(X W^{\top} + B\right),\)
where \(X\) is the input vector, \(W\) and \(B\) are the learnable weight matrix and bias vector, and \(\Phi\) denotes the activation function. Through successive applications of this operation, the network propagates information forward and ultimately produces a prediction for the target variable. During training, the parameters \(W\) and \(B\) are adjusted to minimize the loss between predicted and true stripping times.

For this work, we adopt a fully connected multilayer perceptron (MLP) architecture consisting of one input layer, three hidden layers, and one output layer. Each hidden layer contains 256 neurons and uses the rectified linear unit (ReLU) activation function \citep{Arora16}. This network learns a nonlinear mapping from the six-dimensional phase-space coordinates of a particle to its corresponding stripping time as inferred from N-body simulations.

We train our MLP on the N-body simulations of \citet{Vasiliev21} (hereafter V21). The input features are the six-dimensional phase-space coordinates of each particle, $(\alpha, \delta, \mu_\alpha^*, \mu_\delta, v_{\mathrm{los}}, d)$. The network is designed to predict the stripping time $\hat{T}_s$ while simultaneously outputting the associated log-variance $s = \ln \sigma_{\hat{T}_s}^2$, where taking the logarithm of the variance improves numerical stability during training \citep[see e.g.,][]{Nix94, Kendall17}. Following the convention of \citet{Vasiliev21}, we define stripping time such that negative values indicate the time elapsed since a particle was stripped from the Sgr dSph (e.g., $\hat{T}_s = -1$ Gyr corresponds to a particle stripped 1 Gyr ago), while values equal to zero denote particles still bound to the Sgr dSph. Estimating $s$ is crucial because different tidal wraps overlap in certain regions of phase space, making some particles intrinsically more difficult to assign to a unique stripping epoch. By learning $s$, the network is encouraged to recognize such ambiguous cases and to return appropriately larger predictive uncertainties. We clamp $s$ to the interval $[-9.2,\,2.2]$, which corresponds to confining $\sigma_{\hat{T}_s}$ to $[0.01,\,3]~\mathrm{Gyr}$, thereby excluding unphysical uncertainty estimates and improving numerical stability during optimization.

The loss function adopted in this work is a negative log-likelihood (NLL) loss, defined as  
\begin{equation}\label{eq:loss}
    \mathrm{loss} = \frac{1}{2N} \sum_{i=1}^{N} \left[\, s_i + (T_{s, i} - \hat{T}_{s, i})^2 \exp(-s_i) \right],
\end{equation}
where $\hat{T}_s$ denotes the predicted stripping time and $T_s$ denotes the true stripping time. This formulation corresponds to a heteroscedastic Gaussian likelihood in which the network learns both the mean prediction and its associated uncertainty.

\begin{figure}[htbp]
    \centering
    \includegraphics[width=0.8\linewidth]{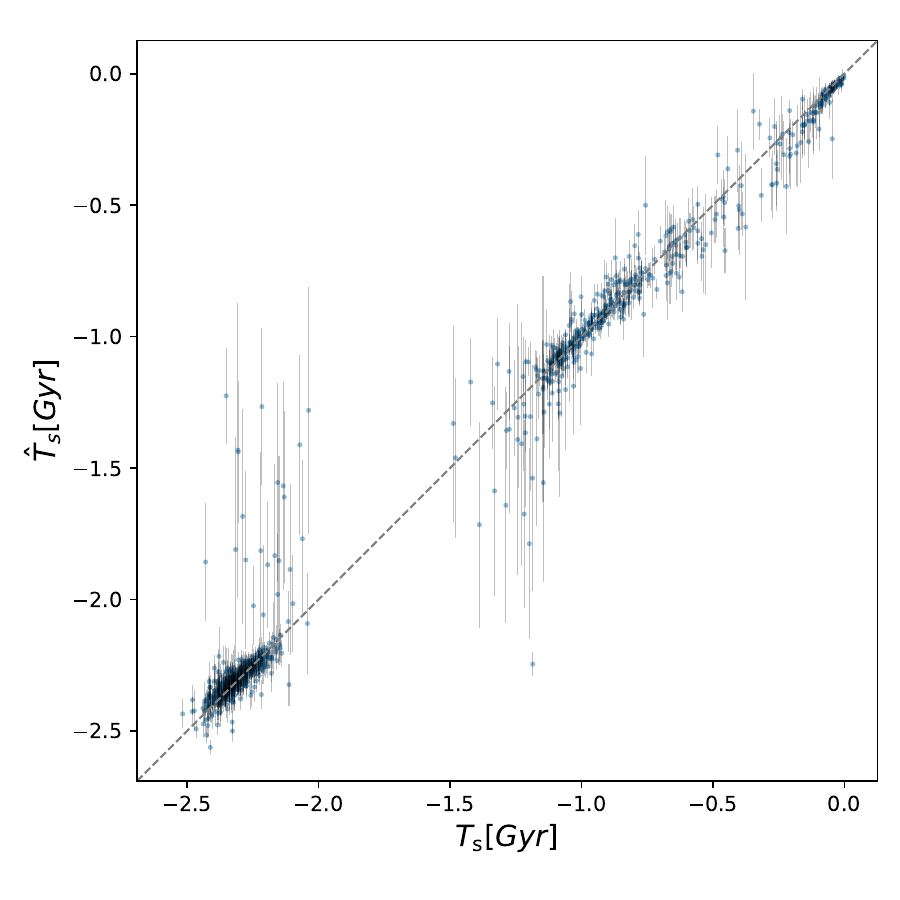}
    \caption{Results on the test set. The x-axis shows the true stripping time $T_s$ of V21 simulation particles, and the y-axis shows the stripping time predicted from the neural network ($\hat{T}_s$). Blue dots with error bars correspond to the adopted value and uncertainty of the NN-predicted results. The grey dashed line corresponds to y=x.}
    \label{fig:mlp_test}
\end{figure}

We selected 5000 particles from the older wrap with $T_s < -2~\mathrm{Gyr}$ and 5000 particles from the younger wrap with $-2~\mathrm{Gyr} \le T_s < 0~\mathrm{Gyr}$. This threshold separates debris stripped during different pericentric passages. Maintaining this balance ensures a well-structured and unbiased training sample, improving the network's ability to distinguish particles originating from different wraps. The synthetic dataset was randomly divided into three subsets: 80\% for training (used to optimize the model parameters), 10\% for validation (used for model selection), and 10\% for testing (used to quantify the generalization error). The model converges after approximately 300 epochs. We select the model with the lowest validation loss and evaluate it on the test set, with results presented in Figure~\ref{fig:mlp_test}.

The predicted stripping times, $\hat{T}_s$, closely track the true values, demonstrating a lack of significant systematic bias. On the test set, the mean squared error (MSE) is $\sim 0.013$, and the mean absolute error (MAE) is $\sim 0.051$. Notably, particles exhibiting larger prediction residuals are consistently assigned larger uncertainties. This indicates that the network not only learns an accurate mapping from phase-space coordinates to stripping time but also effectively identifies regions in phase space where multiple wraps overlap, leading to intrinsically lower predictive certainty. We interpret the network outputs, $\hat{T}_s$ and $\sigma_{\hat{T}_s}^2$, as estimates of the conditional mean $E(T_s|\vec{x}_t) \approx \mu_\theta(\vec{x}_t)$ and variance $\mathrm{Var}(T_s|\vec{x}_t) \approx \sigma^2_\theta(\vec{x}_t)$ of a heteroscedastic Gaussian distribution given true phase-space coordinates $\vec{x}_t$, where $\theta$ represents the network parameters. Upon validating this probabilistic interpretation on the test set, we find that approximately $95\%$ of the true $T_s$ values fall within the predicted $2\sigma$ intervals, confirming that our uncertainty estimates are well-calibrated.

In practice, we only have access to the observed phase-space coordinates, $\vec{x}_o$, and their associated measurement covariance matrix, $\Sigma$. We approximate the posterior distribution of the true coordinates, $\vec{x}_t$, as a multivariate Gaussian: $p(\vec{x}_t|\vec{x}_o) \sim \mathcal{N}(\vec{x}_o, \Sigma)$. To obtain robust predictions, we seek the marginalized mean, $E(T_s|\vec{x}_o)$, and variance, $\mathrm{Var}(T_s|\vec{x}_o)$. These can be derived by marginalizing over the observational uncertainties, utilizing the Law of Total Expectation:

\begin{equation}\label{eq:total_expectation}
\begin{aligned}
E(T_s|\vec{x}_o) 
&= E_{\vec{x}_t|\vec{x}_o} \left[ E(T_s|\vec{x}_t) \right],
\end{aligned}
\end{equation}
where we assume that the stripping time $T_s$ is conditionally independent of the observations $\vec{x}_o$ given the true phase-space coordinates $\vec{x}_t$. Similarly, applying the Law of Total Variance:

\begin{equation}\label{eq:total_variance}
\begin{aligned}
\mathrm{Var}(T_s|\vec{x}_o)
&= E_{\vec{x}_t|\vec{x}_o}\left( \mathrm{Var}(T_s|\vec{x}_t) \right) + \mathrm{Var}_{\vec{x}_t|\vec{x}_o}\left( E(T_s|\vec{x}_t) \right).
\end{aligned}
\end{equation}

Here, $E_{\vec{x}_t|\vec{x}_o}[\cdot]$ and $\mathrm{Var}_{\vec{x}_t|\vec{x}_o}[\cdot]$ denote the expectation and variance taken with respect to the posterior distribution of the true phase-space coordinates, $p(\vec{x}_t|\vec{x}_o)$.
Specifically, for any measurable function $f(\vec{x}_t)$,
\begin{equation}
E_{\vec{x}_t|\vec{x}_o}[\,f(\vec{x}_t)\,] = \int f(\vec{x}_t)\, p(\vec{x}_t|\vec{x}_o) \, d\vec{x}_t ,
\end{equation}
\begin{equation}
\mathrm{Var}_{\vec{x}_t|\vec{x}_o}[\,f(\vec{x}_t)\,] = \int \Big(f(\vec{x}_t)-E_{\vec{x}_t|\vec{x}_o}[f]\Big)^2\, p(\vec{x}_t|\vec{x}_o)\, d\vec{x}_t.
\end{equation}

In practice, we estimate these quantities via MC sampling. We draw $N = 100$ samples $\{\vec{x}_{t, i}\}_{i=1}^{N}$ from the distribution $\mathcal{N}(\vec{x}_o, \Sigma)$ and compute:
\begin{equation}
    E(T_s|\vec{x}_o) \approx \frac{1}{N}\sum_{i=1}^N \mu_\theta (\vec{x}_{t, i}) \equiv \bar{\mu},
\end{equation}
\begin{equation}
    \mathrm{Var}(T_s|\vec{x}_o) \approx \frac{1}{N}\sum_{i=1}^N \sigma^2_\theta (\vec{x}_{t, i}) + \frac{1}{N-1}\sum_{i=1}^N \left( \mu_\theta(\vec{x}_{t, i}) - \bar{\mu} \right)^2,
\end{equation}
where the first term of variance represents the average intrinsic uncertainty from the neural network predictions, and the second term captures the additional uncertainty due to observational errors propagated through the model. This MC marginalization partially mitigates the mismatch between noise-free training features and noisy observations, but it does not eliminate the possibility that measurement errors push some stars outside the feature-space region covered by the training set. We therefore performed a robustness test by retraining the MLP on a mock catalog with observational perturbations added to the six-dimensional input coordinates. The metallicity gradient inferred from this noise-trained model is fully consistent with the fiducial result, indicating that our conclusions are not sensitive to a reasonable level of input noise. Further details are given in Appendix~\ref{sec:noise}.

\section{Results}\label{sec:results}
\subsection{Metallicity Gradient}
Using the marginalization framework described above, we estimated the expected stripping time and its uncertainty for each Sgr Stream star based on its observed phase-space coordinates and covariance. Building on these estimates, we aim to determine the metallicity gradient with respect to stripping time. For the survey metallicities, we converted reported $\mathrm{[Fe/H]}$ values to $\mathrm{[M/H]}_{\mathrm{Survey}}$ using the \citet{Salaris93} transformation when valid $\mathrm{[\alpha/Fe]}$ measurements were available; for stars without reported $\mathrm{[\alpha/Fe]}$, we assumed $\mathrm{[M/H]} \approx \mathrm{[Fe/H]}$. To ensure robust metallicity measurements, we first imposed quality cuts on the metallicity uncertainties from both the original surveys and \texttt{StarHorse}, requiring $\sigma_{\mathrm{Survey}} < 0.5\,\mathrm{dex}$ and $\sigma_{\textsc{sh}} < 0.5\,\mathrm{dex}$. We then filtered for stars where the metallicity derived from \texttt{StarHorse} is consistent with the original survey value within $3\sigma$, ensuring that systematic differences between surveys do not bias our results:
\begin{equation}
    \left| \mathrm{[M/H]}_{\textsc{sh}} - \mathrm{[M/H]}_{\mathrm{Survey}} \right| < 3 \sqrt{\sigma_{\textsc{sh}}^2 + \sigma_{\mathrm{Survey}}^2}.
\end{equation}

\begin{figure}[htbp]
    \centering
    \includegraphics[width=\linewidth]{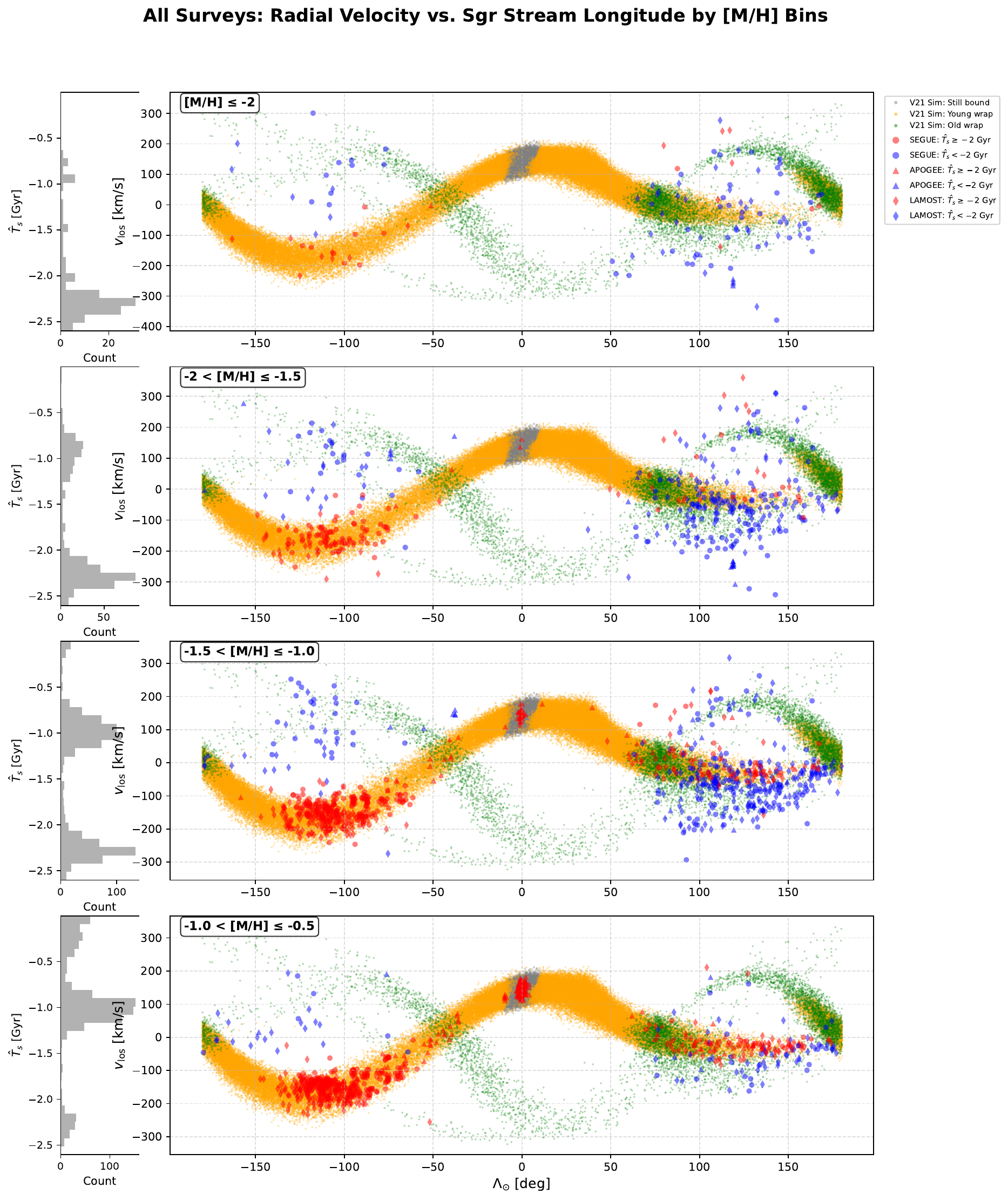}
    \caption{Combined results from SEGUE, APOGEE, and LAMOST surveys. The main panels show the line-of-sight velocity relative to $\Lambda_\odot$ for the Sgr Stream across four metallicity bins. Background gray, orange, and green scatter points represent particles from the V21 simulation classified as still-bound, young wrap (\mbox{$0 >T_{\rm s}\geq -2$~Gyr}), and old wrap (\mbox{$T_{\rm s}< -2$~Gyr}), respectively. Overlaid symbols show observed Sgr Stream members: circles for SEGUE, triangles for APOGEE, and diamonds for LAMOST. Red symbols indicate stars with \mbox{$T_{\rm s}\geq -2$~Gyr}, while blue symbols indicate \mbox{$T_{\rm s}< -2$~Gyr}. The left panels display stripping time distributions for observed stars within each metallicity bin. The figure demonstrates that as metallicity decreases, the fraction of young wraps decreases while old wraps increase.}
    \label{fig:All_surveys_demo}
\end{figure}

We further refined our sample by requiring precise stripping time estimates ($\sigma_{\hat{T}_s} < 0.3\,\mathrm{Gyr}$) and excluding stars with NN-predicted stripping times $\hat{T}_s < -2.607\,\mathrm{Gyr}$ or $\hat{T}_s > 0\,\mathrm{Gyr}$. The latter two cuts correspond to the lower and upper boundaries of stripping times in the \citet{Vasiliev21} simulation; we consider extrapolation beyond this range unreliable for our NN. Applying these criteria yields a final sample of $\sim 1040$ stars in SEGUE, $\sim 570$ in APOGEE DR17, and $\sim 960$ in LAMOST DR7 LRS. Results from all three surveys are presented in Figure~\ref{fig:All_surveys_demo}. In this figure, we show Sgr Stream members in different metallicity bins overlaid on the V21 simulation, with colors indicating stripping time, and display the stripping time distribution in the left panel of each bin.

\begin{figure}[htbp!]
\centering
\begin{tikzpicture}[
    node distance=1.2cm,
    latent/.style={circle,draw,inner sep=1pt,minimum size=18pt},
    obs/.style={circle,draw,fill=gray!20,inner sep=1pt,minimum size=18pt},
    const/.style={rectangle,inner sep=2pt},
    plate/.style={draw,rounded corners=5pt,inner sep=6pt}
]

\node[latent] (w)            at (0,3) {$\mathbf{w}$};
\node[latent] (mu)           at (-2,3) {$\mu_k$};
\node[latent] (tau)          at ( 2,3) {$\tau_k$};

\node[latent] (xstar)        at (0,1.5) {$T_{s,i}$};

\node[latent] (a)            at (-3,1) {$a$};
\node[latent] (b)            at (-3,0) {$b$};
\node[latent] (sint)         at (-3,-1) {$\sigma_{\text{int}}$};

\node[obs] (xobs)            at (-1,-0.5) {$\hat{T}_{s,i}$};
\node[obs] (yobs)            at ( 1,-0.5) {$[M/H]_i$};

\node[const] (ex)            at (-1,-1.7) {$\sigma_{T_{s,i}}$};
\node[const] (ey)            at ( 1,-1.7) {$\sigma_{[M/H]_i}$};

\draw[->] (w)      -- (xstar);
\draw[->] (mu)     -- (xstar);
\draw[->] (tau)    -- (xstar);

\draw[->] (xstar)  -- (xobs);
\draw[->] (xstar)  -- (yobs);

\draw[->] (a)      -- (yobs);
\draw[->] (b)      -- (yobs);
\draw[->] (sint)   -- (yobs);

\draw[->] (ex)     -- (xobs);
\draw[->] (ey)     -- (yobs);

\node[plate,fit=(xstar)(xobs)(yobs)(ex)(ey),
      label=below right:{\small $i=1,\dots,N$}] {};

\node[plate,fit=(mu)(tau),
      label=below right:{\small $k=1,2,3$}] {};

\end{tikzpicture}
\caption{Directed Acyclic Graph (DAG) for our hierarchical Bayesian mixture-plus-regression model. Latent variables are shown as unfilled circles, observed quantities as filled gray circles, and fixed parameters are shown in rectangles. The model infers the true stripping time $T_{s,i}$ from the NN prediction $\hat{T}_{s,i}$ and its uncertainty $\sigma_{T_{s,i}}$. A Gaussian mixture model with three components ($k=1,2,3$), parameterized by weights $\mathbf{w}$, means $\mu_k$, and standard deviations $\tau_k$, describes the distribution of stripping times corresponding to the three main pericentric passages of the Sgr Stream. The metallicity $\mathrm{[M/H]}_{i}$ is modeled as a linear function of the true stripping time with slope $b$, intercept $a$, and intrinsic scatter $\sigma_{\mathrm{int}}$, observed with measurement uncertainty $\sigma_{\mathrm{[M/H]}\,i}$. The plates indicate replication over $N$ stars (bottom) and the three mixture components (top right).}
\label{fig:DAG}
\end{figure}
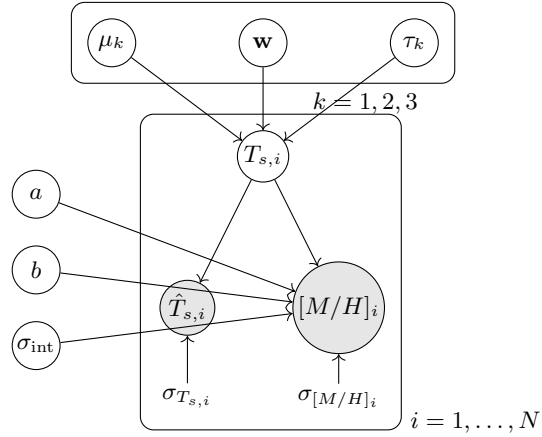

We combined data from the three surveys to derive the metallicity gradient. Given that we identified only $\sim 10$ duplicate pairs between any two surveys, we regarded this overlap as negligible and treated the samples as independent. To simultaneously account for the three distinct pericentric passages and propagate uncertainties from the NN predictions $\hat{T}_s$ to the true stripping times $T_s$, we employed a hierarchical Bayesian mixture-plus-regression framework, as illustrated by the Directed Acyclic Graph (DAG) in Figure~\ref{fig:DAG}.

Our model is physically motivated by the orbital history of the Sgr Stream, which is dominated by three main pericentric passages: the penultimate passage ($T_s < -2\,\mathrm{Gyr}$), the last passage ($-1.5\,\mathrm{Gyr} < T_s < -0.5\,\mathrm{Gyr}$), and the ongoing passage ($T_s > -0.5\,\mathrm{Gyr}$). While the true stripping time, $T_s$, is a latent variable inaccessible to direct observation, we utilize the NN-derived estimates, $\hat{T}_s$, along with their uncertainties, $\sigma_{\hat{T}_s}$ (incorporating both internal NN uncertainty and propagated observational errors). Physically, the stripping time reflects the initial binding energy of the stars within the Sgr progenitor: loosely bound stars, which typically reside in the metal-poor outskirts, are stripped earlier. Consequently, we model the metallicity as a linear function of the stripping time with intrinsic scatter:
\begin{equation}
    \mathrm{[M/H]} \sim \mathcal{N}(b T_s + a, \sigma_{\mathrm{int}}^2),
\end{equation}
where $\sigma_{\mathrm{int}}$ represents the intrinsic dispersion of this relation.

\begin{figure}[htbp]
    \centering
    \includegraphics[width=0.8\linewidth]{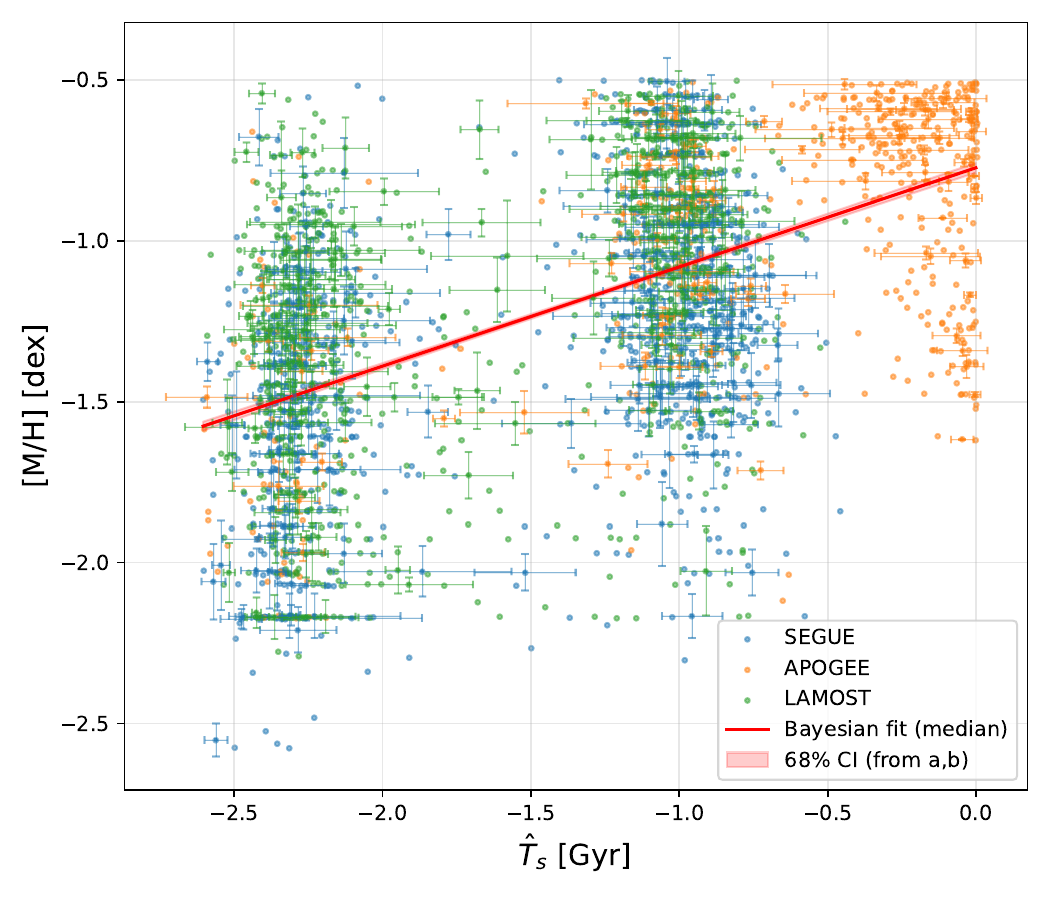}
    \caption{The x-axis shows NN-predicted stripping time $\hat{T}_s$, and the y-axis shows metallicity. Blue, orange, and green dots correspond to data from SEGUE, APOGEE, and LAMOST. The red line shows the metallicity gradient, with the shaded region representing the $1\sigma$ credible interval.}
    \label{fig:met_grad}
\end{figure}

\begin{figure}[htbp]
    \centering
    \includegraphics[width=0.8\linewidth]{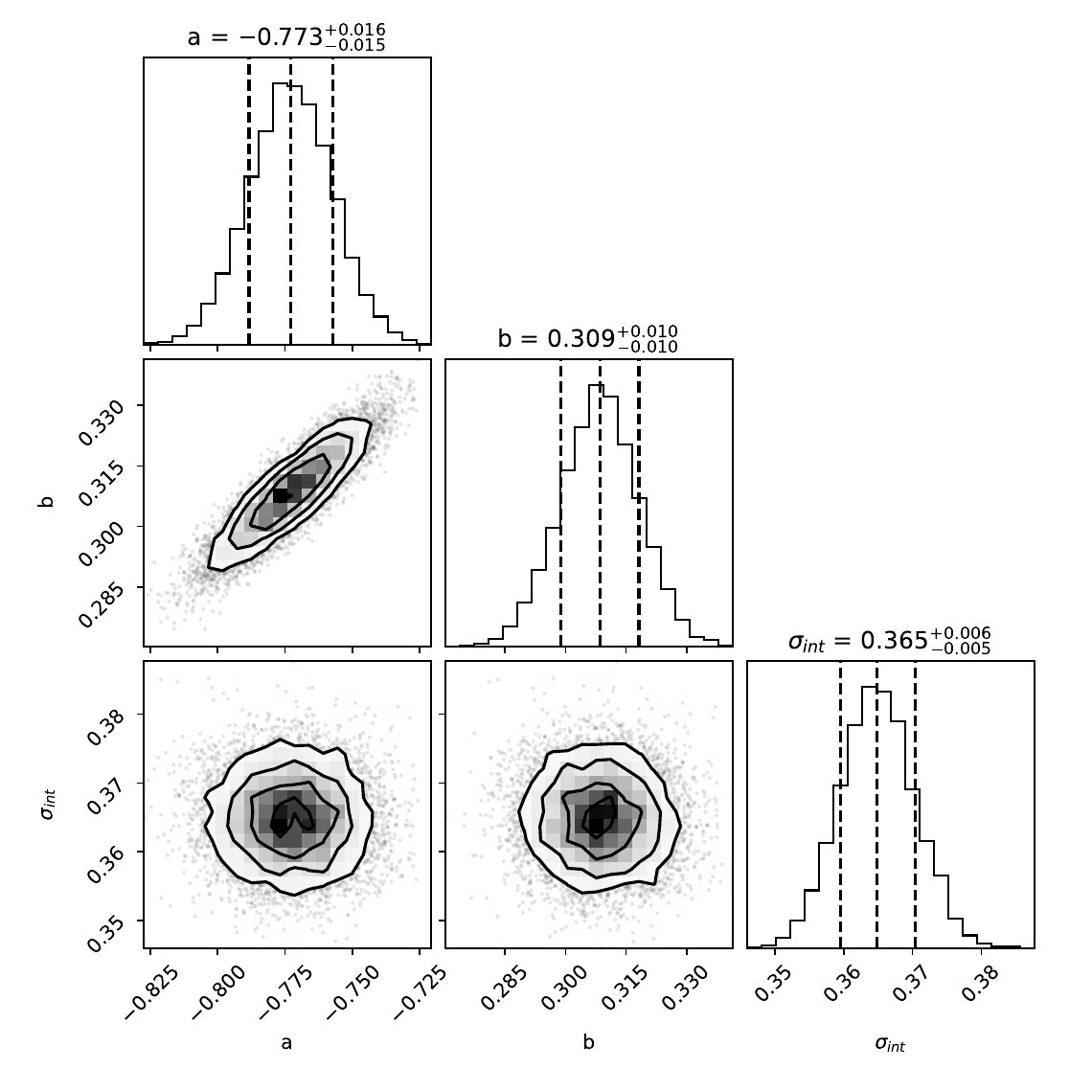}
    \caption{Corner plot of the slope, intercept, and dispersion of the metallicity gradient.}
    \label{fig:corner_plot}
\end{figure}

Figure~\ref{fig:met_grad} displays the metallicity distribution ($[\mathrm{M/H}]$) as a function of NN-predicted stripping time ($\hat{T}_s$), with data from SEGUE, APOGEE, and LAMOST represented by blue, orange, and green points, respectively. For visual clarity and to avoid overcrowding, we display error bars for only a random 15\% subsample of the data. The inferred metallicity gradient and its associated $1\sigma$ credible interval are overlaid as a red line and shaded region. The posterior distributions of the model parameters $a$, $b$, and $\sigma_{\mathrm{int}}$ are presented in the corner plot (Figure~\ref{fig:corner_plot}), demonstrating excellent convergence. The best-fit parameters are slope $b=0.309 \pm 0.010\,\mathrm{dex\,Gyr^{-1}}$, intercept $a=-0.773 \pm 0.016\,\mathrm{dex}$, and intrinsic scatter $\sigma_{\mathrm{int}} = 0.365 \pm 0.006\,\mathrm{dex}$. We find consistent metallicity gradients across all three surveys. For example, SEGUE and LAMOST, which contain primarily Sgr Stream members, yield gradients similar to APOGEE, which contains many stars in the Sgr dSph core, and all three values agree with our combined result. Our gradient is approximately two times steeper than that of \citet{Hayes20}, who reported a metallicity gradient of $0.12 \pm 0.03$ dex\,Gyr$^{-1}$. \citet{Hayes20} estimated stripping times by computing a distance metric between observed stars and particles in simulations of \citet{Law10} and then assigning $T_s$ from the nearby simulation particles' average $T_s$ if the distance lies below an ad hoc threshold. However, such heuristic methods can introduce biases, especially in regions where multiple wraps overlap and the nearby-particle assignment becomes unreliable. Our NN framework provides a more flexible, data-driven alternative that naturally accounts for these ambiguities through its predicted uncertainties. Applying our method to the \citet{Hayes20} Sgr Stream dataset yields a metallicity gradient of $\sim 0.3$ dex\,Gyr$^{-1}$, which is consistent with our result considering the much smaller sample size of \citet{Hayes20} (166 Sgr Stream stars).

\subsection{Stripping time of globular clusters}

Several globular clusters (GCs) have been suggested to originate from the Sgr dSph \citep{Sohn18, Tang18, Massari19, Bellazzini20}. In this work, we primarily adopt the sample defined by \citet{Bellazzini20}, who investigated the association of Galactic GCs with the Sgr Stream using \textit{Gaia} DR2 data. By utilizing RR Lyrae variables to trace the Sgr Stream in 6D phase space, they identified GCs spatially and kinematically coincident with the Sgr Stream. Beyond the GCs residing in the main body of the Sgr dSph (M\,54, Ter\,8, Ter\,7, Arp\,2), they confirmed the membership of Pal\,12 and Whiting\,1 in the trailing arm, which is populated by stars lost during recent perigalactic passages. Additionally, they highlighted NGC\,2419, NGC\,5634, and NGC\,4147 as intriguing candidates, possibly associated with more ancient wraps of the Sgr Stream.

Applying the same Sgr Stream selection criteria as \citet{Limberg23}, we identify the same set of seven GCs (M\,54, Whiting\,1, Koposov\,1, Terzan\,7, Arp\,2, Terzan\,8, and Pal\,12). We refine this sample by excluding Koposov\,1, for which there is limited literature support for association with the Sgr dSph or the Sgr Stream. We additionally include NGC\,2419, NGC\,5634, and NGC\,4147 as literature-motivated candidates. However, these GCs do not satisfy the Sgr Stream selection criteria adopted here. In particular, the membership of NGC\,5634 and NGC\,4147 in the Sgr system remains uncertain, so our analysis does not seek to establish their membership; rather, it estimates the stripping times implied if they are eventually confirmed to be associated with the Sgr system.

We adopted mean proper motions and distances for the GCs from \citet{Vasiliev&Baumgardt2021} and \citet{Baumgardt21}, respectively. By inputting the phase-space coordinates and covariance matrices of these GCs into our NN and applying the marginalization scheme described previously, we derived their stripping times. The detailed results are presented in Table~\ref{tab:sgr_gc_ts}.

As expected, M\,54, Ter\,7, Ter\,8, and Arp\,2 exhibit stripping times consistent with zero within uncertainties, confirming that they are currently bound to the Sgr dSph. For the detached GCs with stronger evidence for Sgr association, we infer stripping times of $-0.9 \pm 0.1$\,Gyr for Pal\,12, $-1.1 \pm 0.2$\,Gyr for Whiting\,1, and $-2.1 \pm 0.2$\,Gyr for NGC\,2419. For NGC\,4147 and NGC\,5634, whose membership remains debated, our model yields stripping times of $-1.1 \pm 0.4$\,Gyr and $-1.1 \pm 0.1$\,Gyr, respectively; these values should be interpreted conditionally, namely as the stripping epochs implied if they are genuine members of the Sgr system. Notably, our finding that NGC\,2419 was stripped $\sim 2.1$\,Gyr ago is consistent with previous suggestions \citep[e.g.,][]{Massari17, Belokurov14, Bellazzini20} that it is associated with a distant and ancient arm of the Sgr Stream.

\begin{table}[htbp!]
\centering
\caption{Predicted stripping times for GCs associated with the Sgr dSph}
\label{tab:sgr_gc_ts}
\begin{tabular}{lcc}
\hline\hline
Cluster & $\hat{T}_s$ (Gyr) & $\sigma_{\hat{T}_s}$ (Gyr) \\
\hline
\multicolumn{3}{l}{\textit{Still bound}} \\
M54 (NGC 6715) &  0.0 & 0.02 \\
Ter 7          & -0.2 & 0.2 \\
Arp 2          & -0.02 & 0.01 \\
Ter 8          & -0.03 & 0.03 \\
\hline
\multicolumn{3}{l}{\textit{Stripped}} \\
Pal 12         & -0.9 & 0.1 \\
Whiting 1      & -1.1 & 0.2 \\
NGC 4147       & -1.1 & 0.4 \\
NGC 5634       & -1.1 & 0.1 \\
NGC 2419       & -2.1 & 0.2 \\
\hline
\end{tabular}
\end{table}

\section{Discussion}\label{sec:discussion}

In this work, we have measured the metallicity gradient as a function of stripping time. Since the stripping time serves as a proxy for the initial binding energy of the stars, this approach enables us to effectively utilize data that lacks a distinct spatial metallicity gradient—due to phase mixing—to place robust constraints on chemo-dynamical models of the Sgr dSph. Future work could extend this framework to directly quantify the metallicity gradient with respect to binding energy, providing further insights into the chemical evolution of the Sgr progenitor.

We acknowledge two main limitations in this work. First, our NN was trained exclusively on the N-body simulation of \citet{Vasiliev21}. Different simulations employ distinct potential models and initial conditions, and thus cannot be treated as samples drawn from the same underlying distribution. Consequently, generalizing a NN trained on one simulation to another is non-trivial. To quantify this limitation, we tested our V21-trained NN on the simulation from \citet{Wang22}. For the young wrap (recent debris), the predicted stripping times agree reasonably well with the true values, albeit with increased scatter. In contrast, for the old wrap (ancient debris), the predictions exhibit a significant systematic bias. This performance difference arises because both simulations were calibrated using the same observational constraints from \citet{Vasiliev21}, which primarily constrain the prominent young wraps of the Sgr Stream. Consequently, while the young wraps remain structurally similar across models, the older, less constrained wraps diverge significantly, limiting the NN's cross-simulation accuracy in those regions.

A second limitation arises from the nature of our training data: our model was trained on a pure Sgr Stream simulation without Milky Way contamination, whereas observational datasets inevitably contain foreground and background stars. Although the selection criteria we applied in Section~\ref{subsec:select}—based on \citet{Limberg23}—are stringent and expected to exclude $\sim 90\%$ of potential Gaia-Sausage-Enceladus (GSE) stars, residual contamination cannot be completely ruled out. A more robust approach would involve training the NN to explicitly distinguish true Sgr Stream members from field contaminants. However, incorporating such membership classification is beyond the scope of the present study and will be addressed in future work.

\section{Conclusion}\label{sec:conclusion}

In this study, we developed a neural-network-based framework trained on $N$-body simulations to reconstruct the stripping history of the Sgr Stream directly from 6D phase-space coordinates. By applying this method to a comprehensive sample of stars combining \textit{Gaia} EDR3 astrometry with spectroscopy from SEGUE, APOGEE, and LAMOST, we arrived at the following main conclusions:

\begin{enumerate}
    \item We detected a significant positive correlation between metallicity and stripping time, characterized by a linear gradient of $\sim 0.3\,\mathrm{dex\,Gyr^{-1}}$. This result confirms that stars stripped earlier—originating from the loosely bound, metal-poor outskirts of the progenitor—are systematically more metal-poor than those stripped recently. This demonstrates that stripping time serves as an effective dynamical proxy for binding energy, allowing us to recover intrinsic chemo-dynamical trends even in regions where spatial phase mixing obscures traditional gradients.
    \item We provided quantitative stripping time estimates for GCs associated with or proposed to belong to the Sgr dSph or the Sgr Stream. Our results confirm that GCs such as M\,54 and Ter\,7 remain currently bound to the Sgr dSph, and indicate that NGC\,2419 was likely stripped during an earlier passage. Notably, the inferred stripping time of $\sim 2.1\,\mathrm{Gyr}$ ago for NGC\,2419 supports the hypothesis that it is associated with an ancient wrap of the Sgr Stream.
    \item Our work highlights the power of data-driven models in disentangling complex tidal debris. By mapping phase-space coordinates to stripping times, we offer a robust pathway to constrain the accretion history and chemical evolution of dwarf galaxy progenitors.
\end{enumerate}

\begin{acknowledgments}
This work was supported by National Key R\&D Program of China No. 2024YFA1611900, and the National Natural Science Foundation of China (NSFC Nos. 11973042, 11973052). 

This work has made use of data from the European Space Agency (ESA) mission Gaia (\url{https://www.cosmos.esa.int/gaia}), processed by the Gaia Data Processing and Analysis Consortium (DPAC, \url{https://www.cosmos.esa.int/web/gaia/dpac/consortium}). Funding for the DPAC has been provided by national institutions, in particular the institutions participating in the Gaia Multilateral Agreement. 

Funding for the Sloan Digital Sky Survey IV has been provided by the Alfred P. Sloan Foundation, the U.S. Department of Energy Office of Science, and the Participating Institutions. SDSS-IV acknowledges support and resources from the Center for High Performance Computing at the University of Utah. The SDSS website is www.sdss.org. SDSS-IV is managed by the Astrophysical Research Consortium for the Participating Institutions of the SDSS Collaboration.

LAMOST is a National Major Scientific Project built by the Chinese Academy of Sciences, funded by the National Development and Reform Commission.
\end{acknowledgments}

\software{astropy \citep{Astropy13, Astropy18, Astropy22}, Agama \citep{Vasiliev19}, pytorch \citep{Paszke19}}

\appendix
\section{Robustness to Observational Noise}\label{sec:noise}

To test whether the mismatch between noise-free training inputs and noisy observations could bias our conclusions, we retrained the MLP on a mock catalog with observational perturbations applied to the six-dimensional phase-space inputs. The goal of this exercise is not to build a fully realistic forward model, but to verify that our inferred metallicity gradient remains stable under a reasonable level of input noise.

We construct the noisy mock catalog from the V21 simulation using \texttt{py-ananke} \citep{Thob24} to generate Gaia-like astrometry and photometry. Because our observational analysis is based on \textit{Gaia} EDR3, we rescale the \texttt{py-ananke} astrometric uncertainties from their DR2-like values using approximate DR2-to-DR3 improvement factors \citep{Gaia21}, and we retain only stars with $G<19$ to match the observational sample. We then perturb the remaining observables with simple empirical uncertainty models: fractional distance errors are drawn uniformly over the range relevant to our sample, radial-velocity uncertainties are assigned as functions of Gaia $G$ magnitude, and each mock star is probabilistically associated with SEGUE, APOGEE, or LAMOST based on its location in observable space. This simplified prescription is intended only to approximately reproduce the scale of the observational uncertainties.

Using this noise-added synthetic catalog, we repeat the same training and test procedure as for the fiducial noise-free model. On the synthetic test set, the noise-trained network shows the expected degradation in predictive accuracy, with MSE $\sim 0.023$ and MAE $\sim 0.087$, compared with $\sim 0.013$ and $\sim 0.051$ for the noise-free model. Nevertheless, the predicted stripping times remain in good overall agreement with the true values, and the network continues to assign larger uncertainties in regions where the phase-space mapping is less reliable.

We then apply the noise-trained network to the observational sample and repeat the same MC marginalization and metallicity-gradient fit used in the main analysis. After the same quality cuts, the final sample contains $\sim 1010$ SEGUE stars, $\sim 370$ APOGEE stars, and $\sim 820$ LAMOST stars. The inferred gradient is $b=0.313 \pm 0.011\,\mathrm{dex\,Gyr^{-1}}$, with intercept $a=-0.801 \pm 0.018\,\mathrm{dex}$ and intrinsic scatter $\sigma_{\mathrm{int}} = 0.374 \pm 0.006\,\mathrm{dex}$. These values are fully consistent, within the uncertainties, with the fiducial results in the main text. We therefore conclude that although input noise modestly degrades the NN predictive accuracy, it does not materially affect the inferred metallicity gradient, demonstrating that our main conclusion is robust to reasonable perturbations of the 6D inputs.

\bibliography{myBib}{}

@ARTICLE{Astropy22,
       author = {{Astropy Collaboration} and {Price-Whelan}, Adrian M. and {Lim}, Pey Lian and {Earl}, Nicholas and {Starkman}, Nathaniel and {Bradley}, Larry and {Shupe}, David L. and {Patil}, Aarya A. and {Corrales}, Lia and {Brasseur}, C.~E. and {N{\"o}the}, Maximilian and {Donath}, Axel and {Tollerud}, Erik and {Morris}, Brett M. and {Ginsburg}, Adam and {Vaher}, Eero and {Weaver}, Benjamin A. and {Tocknell}, James and {Jamieson}, William and {van Kerkwijk}, Marten H. and {Robitaille}, Thomas P. and {Merry}, Bruce and {Bachetti}, Matteo and {G{\"u}nther}, H. Moritz and {Aldcroft}, Thomas L. and {Alvarado-Montes}, Jaime A. and {Archibald}, Anne M. and {B{\'o}di}, Attila and {Bapat}, Shreyas and {Barentsen}, Geert and {Baz{\'a}n}, Juanjo and {Biswas}, Manish and {Boquien}, M{\'e}d{\'e}ric and {Burke}, D.~J. and {Cara}, Daria and {Cara}, Mihai and {Conroy}, Kyle E. and {Conseil}, Simon and {Craig}, Matthew W. and {Cross}, Robert M. and {Cruz}, Kelle L. and {D'Eugenio}, Francesco and {Dencheva}, Nadia and {Devillepoix}, Hadrien A.~R. and {Dietrich}, J{\"o}rg P. and {Eigenbrot}, Arthur Davis and {Erben}, Thomas and {Ferreira}, Leonardo and {Foreman-Mackey}, Daniel and {Fox}, Ryan and {Freij}, Nabil and {Garg}, Suyog and {Geda}, Robel and {Glattly}, Lauren and {Gondhalekar}, Yash and {Gordon}, Karl D. and {Grant}, David and {Greenfield}, Perry and {Groener}, Austen M. and {Guest}, Steve and {Gurovich}, Sebastian and {Handberg}, Rasmus and {Hart}, Akeem and {Hatfield-Dodds}, Zac and {Homeier}, Derek and {Hosseinzadeh}, Griffin and {Jenness}, Tim and {Jones}, Craig K. and {Joseph}, Prajwel and {Kalmbach}, J. Bryce and {Karamehmetoglu}, Emir and {Ka{\l}uszy{\'n}ski}, Miko{\l}aj and {Kelley}, Michael S.~P. and {Kern}, Nicholas and {Kerzendorf}, Wolfgang E. and {Koch}, Eric W. and {Kulumani}, Shankar and {Lee}, Antony and {Ly}, Chun and {Ma}, Zhiyuan and {MacBride}, Conor and {Maljaars}, Jakob M. and {Muna}, Demitri and {Murphy}, N.~A. and {Norman}, Henrik and {O'Steen}, Richard and {Oman}, Kyle A. and {Pacifici}, Camilla and {Pascual}, Sergio and {Pascual-Granado}, J. and {Patil}, Rohit R. and {Perren}, Gabriel I. and {Pickering}, Timothy E. and {Rastogi}, Tanuj and {Roulston}, Benjamin R. and {Ryan}, Daniel F. and {Rykoff}, Eli S. and {Sabater}, Jose and {Sakurikar}, Parikshit and {Salgado}, Jes{\'u}s and {Sanghi}, Aniket and {Saunders}, Nicholas and {Savchenko}, Volodymyr and {Schwardt}, Ludwig and {Seifert-Eckert}, Michael and {Shih}, Albert Y. and {Jain}, Anany Shrey and {Shukla}, Gyanendra and {Sick}, Jonathan and {Simpson}, Chris and {Singanamalla}, Sudheesh and {Singer}, Leo P. and {Singhal}, Jaladh and {Sinha}, Manodeep and {Sip{\H{o}}cz}, Brigitta M. and {Spitler}, Lee R. and {Stansby}, David and {Streicher}, Ole and {{\v{S}}umak}, Jani and {Swinbank}, John D. and {Taranu}, Dan S. and {Tewary}, Nikita and {Tremblay}, Grant R. and {de Val-Borro}, Miguel and {Van Kooten}, Samuel J. and {Vasovi{\'c}}, Zlatan and {Verma}, Shresth and {de Miranda Cardoso}, Jos{\'e} Vin{\'\i}cius and {Williams}, Peter K.~G. and {Wilson}, Tom J. and {Winkel}, Benjamin and {Wood-Vasey}, W.~M. and {Xue}, Rui and {Yoachim}, Peter and {Zhang}, Chen and {Zonca}, Andrea and {Astropy Project Contributors}},
        title = "{The Astropy Project: Sustaining and Growing a Community-oriented Open-source Project and the Latest Major Release (v5.0) of the Core Package}",
      journal = {\apj},
         year = 2022,
        month = aug,
       volume = {935},
       number = {2},
          eid = {167},
        pages = {167},
          doi = {10.3847/1538-4357/ac7c74},
archivePrefix = {arXiv},
       eprint = {2206.14220},
 primaryClass = {astro-ph.IM},
       adsurl = {https://ui.adsabs.harvard.edu/abs/2022ApJ...935..167A}
}

@ARTICLE{Astropy18,
       author = {{Astropy Collaboration} and {Price-Whelan}, A.~M. and {Sip{\H{o}}cz}, B.~M. and {G{\"u}nther}, H.~M. and {Lim}, P.~L. and {Crawford}, S.~M. and {Conseil}, S. and {Shupe}, D.~L. and {Craig}, M.~W. and {Dencheva}, N. and {Ginsburg}, A. and {VanderPlas}, J.~T. and {Bradley}, L.~D. and {P{\'e}rez-Su{\'a}rez}, D. and {de Val-Borro}, M. and {Aldcroft}, T.~L. and {Cruz}, K.~L. and {Robitaille}, T.~P. and {Tollerud}, E.~J. and {Ardelean}, C. and {Babej}, T. and {Bach}, Y.~P. and {Bachetti}, M. and {Bakanov}, A.~V. and {Bamford}, S.~P. and {Barentsen}, G. and {Barmby}, P. and {Baumbach}, A. and {Berry}, K.~L. and {Biscani}, F. and {Boquien}, M. and {Bostroem}, K.~A. and {Bouma}, L.~G. and {Brammer}, G.~B. and {Bray}, E.~M. and {Breytenbach}, H. and {Buddelmeijer}, H. and {Burke}, D.~J. and {Calderone}, G. and {Cano Rodr{\'\i}guez}, J.~L. and {Cara}, M. and {Cardoso}, J.~V.~M. and {Cheedella}, S. and {Copin}, Y. and {Corrales}, L. and {Crichton}, D. and {D'Avella}, D. and {Deil}, C. and {Depagne}, {\'E}. and {Dietrich}, J.~P. and {Donath}, A. and {Droettboom}, M. and {Earl}, N. and {Erben}, T. and {Fabbro}, S. and {Ferreira}, L.~A. and {Finethy}, T. and {Fox}, R.~T. and {Garrison}, L.~H. and {Gibbons}, S.~L.~J. and {Goldstein}, D.~A. and {Gommers}, R. and {Greco}, J.~P. and {Greenfield}, P. and {Groener}, A.~M. and {Grollier}, F. and {Hagen}, A. and {Hirst}, P. and {Homeier}, D. and {Horton}, A.~J. and {Hosseinzadeh}, G. and {Hu}, L. and {Hunkeler}, J.~S. and {Ivezi{\'c}}, {\v{Z}}. and {Jain}, A. and {Jenness}, T. and {Kanarek}, G. and {Kendrew}, S. and {Kern}, N.~S. and {Kerzendorf}, W.~E. and {Khvalko}, A. and {King}, J. and {Kirkby}, D. and {Kulkarni}, A.~M. and {Kumar}, A. and {Lee}, A. and {Lenz}, D. and {Littlefair}, S.~P. and {Ma}, Z. and {Macleod}, D.~M. and {Mastropietro}, M. and {McCully}, C. and {Montagnac}, S. and {Morris}, B.~M. and {Mueller}, M. and {Mumford}, S.~J. and {Muna}, D. and {Murphy}, N.~A. and {Nelson}, S. and {Nguyen}, G.~H. and {Ninan}, J.~P. and {N{\"o}the}, M. and {Ogaz}, S. and {Oh}, S. and {Parejko}, J.~K. and {Parley}, N. and {Pascual}, S. and {Patil}, R. and {Patil}, A.~A. and {Plunkett}, A.~L. and {Prochaska}, J.~X. and {Rastogi}, T. and {Reddy Janga}, V. and {Sabater}, J. and {Sakurikar}, P. and {Seifert}, M. and {Sherbert}, L.~E. and {Sherwood-Taylor}, H. and {Shih}, A.~Y. and {Sick}, J. and {Silbiger}, M.~T. and {Singanamalla}, S. and {Singer}, L.~P. and {Sladen}, P.~H. and {Sooley}, K.~A. and {Sornarajah}, S. and {Streicher}, O. and {Teuben}, P. and {Thomas}, S.~W. and {Tremblay}, G.~R. and {Turner}, J.~E.~H. and {Terr{\'o}n}, V. and {van Kerkwijk}, M.~H. and {de la Vega}, A. and {Watkins}, L.~L. and {Weaver}, B.~A. and {Whitmore}, J.~B. and {Woillez}, J. and {Zabalza}, V. and {Astropy Contributors}},
        title = "{The Astropy Project: Building an Open-science Project and Status of the v2.0 Core Package}",
      journal = {\aj},
         year = 2018,
        month = sep,
       volume = {156},
       number = {3},
          eid = {123},
        pages = {123},
          doi = {10.3847/1538-3881/aabc4f},
archivePrefix = {arXiv},
       eprint = {1801.02634},
 primaryClass = {astro-ph.IM},
       adsurl = {https://ui.adsabs.harvard.edu/abs/2018AJ....156..123A}
}

@ARTICLE{Astropy13,
       author = {{Astropy Collaboration} and {Robitaille}, Thomas P. and
         {Tollerud}, Erik J. and {Greenfield}, Perry and {Droettboom}, Michael and
         {Bray}, Erik and {Aldcroft}, Tom and {Davis}, Matt and
         {Ginsburg}, Adam and {Price-Whelan}, Adrian M. and
         {Kerzendorf}, Wolfgang E. and {Conley}, Alexander and {Crighton}, Neil and
         {Barbary}, Kyle and {Muna}, Demitri and {Ferguson}, Henry and
         {Grollier}, Fr{\'e}d{\'e}ric and {Parikh}, Madhura M. and
         {Nair}, Prasanth H. and {Unther}, Hans M. and {Deil}, Christoph and
         {Woillez}, Julien and {Conseil}, Simon and {Kramer}, Roban and
         {Turner}, James E.~H. and {Singer}, Leo and {Fox}, Ryan and
         {Weaver}, Benjamin A. and {Zabalza}, Victor and {Edwards}, Zachary I. and
         {Azalee Bostroem}, K. and {Burke}, D.~J. and {Casey}, Andrew R. and
         {Crawford}, Steven M. and {Dencheva}, Nadia and {Ely}, Justin and
         {Jenness}, Tim and {Labrie}, Kathleen and {Lim}, Pey Lian and
         {Pierfederici}, Francesco and {Pontzen}, Andrew and {Ptak}, Andy and
         {Refsdal}, Brian and {Servillat}, Mathieu and {Streicher}, Ole},
        title = "{Astropy: A community Python package for astronomy}",
      journal = {\aap},
         year = "2013",
        month = "Oct",
       volume = {558},
          eid = {A33},
        pages = {A33},
          doi = {10.1051/0004-6361/201322068},
archivePrefix = {arXiv},
       eprint = {1307.6212},
 primaryClass = {astro-ph.IM},
       adsurl = {https://ui.adsabs.harvard.edu/abs/2013A&A...558A..33A}
}

@ARTICLE{White78,
       author = {{White}, S.~D.~M. and {Rees}, M.~J.},
        title = "{Core condensation in heavy halos: a two-stage theory for galaxy formation and clustering.}",
      journal = {\mnras},
         year = 1978,
        month = may,
       volume = {183},
        pages = {341-358},
          doi = {10.1093/mnras/183.3.341},
       adsurl = {https://ui.adsabs.harvard.edu/abs/1978MNRAS.183..341W}
}

@ARTICLE{Springel05,
       author = {{Springel}, Volker and {White}, Simon D.~M. and {Jenkins}, Adrian and {Frenk}, Carlos S. and {Yoshida}, Naoki and {Gao}, Liang and {Navarro}, Julio and {Thacker}, Robert and {Croton}, Darren and {Helly}, John and {Peacock}, John A. and {Cole}, Shaun and {Thomas}, Peter and {Couchman}, Hugh and {Evrard}, August and {Colberg}, J{\"o}rg and {Pearce}, Frazer},
        title = "{Simulations of the formation, evolution and clustering of galaxies and quasars}",
      journal = {\nat},
         year = 2005,
        month = jun,
       volume = {435},
       number = {7042},
        pages = {629-636},
          doi = {10.1038/nature03597},
archivePrefix = {arXiv},
       eprint = {astro-ph/0504097},
 primaryClass = {astro-ph},
       adsurl = {https://ui.adsabs.harvard.edu/abs/2005Natur.435..629S}
}

@ARTICLE{Majewski03,
       author = {{Majewski}, Steven R. and {Skrutskie}, M.~F. and {Weinberg}, Martin D. and {Ostheimer}, James C.},
        title = "{A Two Micron All Sky Survey View of the Sagittarius Dwarf Galaxy. I. Morphology of the Sagittarius Core and Tidal Arms}",
      journal = {\apj},
         year = 2003,
        month = dec,
       volume = {599},
       number = {2},
        pages = {1082-1115},
          doi = {10.1086/379504},
archivePrefix = {arXiv},
       eprint = {astro-ph/0304198},
 primaryClass = {astro-ph},
       adsurl = {https://ui.adsabs.harvard.edu/abs/2003ApJ...599.1082M}
}

@ARTICLE{Ibata94,
       author = {{Ibata}, R.~A. and {Gilmore}, G. and {Irwin}, M.~J.},
        title = "{A dwarf satellite galaxy in Sagittarius}",
      journal = {\nat},
         year = 1994,
        month = jul,
       volume = {370},
       number = {6486},
        pages = {194-196},
          doi = {10.1038/370194a0},
       adsurl = {https://ui.adsabs.harvard.edu/abs/1994Natur.370..194I}
}

@ARTICLE{Bellazzini20,
       author = {{Bellazzini}, M. and {Ibata}, R. and {Malhan}, K. and {Martin}, N. and {Famaey}, B. and {Thomas}, G.},
        title = "{Globular clusters in the Sagittarius stream. Revising members and candidates with Gaia DR2}",
      journal = {\aap},
         year = 2020,
        month = apr,
       volume = {636},
          eid = {A107},
        pages = {A107},
          doi = {10.1051/0004-6361/202037621},
archivePrefix = {arXiv},
       eprint = {2003.07871},
 primaryClass = {astro-ph.GA},
       adsurl = {https://ui.adsabs.harvard.edu/abs/2020A&A...636A.107B}
}

@ARTICLE{Vasiliev21,
       author = {{Vasiliev}, Eugene and {Belokurov}, Vasily and {Erkal}, Denis},
        title = "{Tango for three: Sagittarius, LMC, and the Milky Way}",
      journal = {\mnras},
         year = 2021,
        month = feb,
       volume = {501},
       number = {2},
        pages = {2279-2304},
          doi = {10.1093/mnras/staa3673},
archivePrefix = {arXiv},
       eprint = {2009.10726},
 primaryClass = {astro-ph.GA},
       adsurl = {https://ui.adsabs.harvard.edu/abs/2021MNRAS.501.2279V}
}

@ARTICLE{Helmi99,
       author = {{Helmi}, Amina and {White}, Simon D.~M.},
        title = "{Building up the stellar halo of the Galaxy}",
      journal = {\mnras},
         year = 1999,
        month = aug,
       volume = {307},
       number = {3},
        pages = {495-517},
          doi = {10.1046/j.1365-8711.1999.02616.x},
archivePrefix = {arXiv},
       eprint = {astro-ph/9901102},
 primaryClass = {astro-ph},
       adsurl = {https://ui.adsabs.harvard.edu/abs/1999MNRAS.307..495H}
}

@ARTICLE{Belokurov18,
       author = {{Belokurov}, V. and {Erkal}, D. and {Evans}, N.~W. and {Koposov}, S.~E. and {Deason}, A.~J.},
        title = "{Co-formation of the disc and the stellar halo}",
      journal = {\mnras},
         year = 2018,
        month = jul,
       volume = {478},
       number = {1},
        pages = {611-619},
          doi = {10.1093/mnras/sty982},
archivePrefix = {arXiv},
       eprint = {1802.03414},
 primaryClass = {astro-ph.GA},
       adsurl = {https://ui.adsabs.harvard.edu/abs/2018MNRAS.478..611B}
}

@ARTICLE{Helmi18,
       author = {{Helmi}, Amina and {Babusiaux}, Carine and {Koppelman}, Helmer H. and {Massari}, Davide and {Veljanoski}, Jovan and {Brown}, Anthony G.~A.},
        title = "{The merger that led to the formation of the Milky Way's inner stellar halo and thick disk}",
      journal = {\nat},
         year = 2018,
        month = oct,
       volume = {563},
       number = {7729},
        pages = {85-88},
          doi = {10.1038/s41586-018-0625-x},
archivePrefix = {arXiv},
       eprint = {1806.06038},
 primaryClass = {astro-ph.GA},
       adsurl = {https://ui.adsabs.harvard.edu/abs/2018Natur.563...85H}
}

@ARTICLE{Koppelman19,
       author = {{Koppelman}, Helmer H. and {Helmi}, Amina and {Massari}, Davide and {Price-Whelan}, Adrian M. and {Starkenburg}, Tjitske K.},
        title = "{Multiple retrograde substructures in the Galactic halo: A shattered view of Galactic history}",
      journal = {\aap},
         year = 2019,
        month = nov,
       volume = {631},
          eid = {L9},
        pages = {L9},
          doi = {10.1051/0004-6361/201936738},
archivePrefix = {arXiv},
       eprint = {1909.08924},
 primaryClass = {astro-ph.GA},
       adsurl = {https://ui.adsabs.harvard.edu/abs/2019A&A...631L...9K}
}

@ARTICLE{Myeong19,
       author = {{Myeong}, G.~C. and {Vasiliev}, E. and {Iorio}, G. and {Evans}, N.~W. and {Belokurov}, V.},
        title = "{Evidence for two early accretion events that built the Milky Way stellar halo}",
      journal = {\mnras},
         year = 2019,
        month = sep,
       volume = {488},
       number = {1},
        pages = {1235-1247},
          doi = {10.1093/mnras/stz1770},
archivePrefix = {arXiv},
       eprint = {1904.03185},
 primaryClass = {astro-ph.GA},
       adsurl = {https://ui.adsabs.harvard.edu/abs/2019MNRAS.488.1235M}
}

@ARTICLE{Naidu20,
       author = {{Naidu}, Rohan P. and {Conroy}, Charlie and {Bonaca}, Ana and {Johnson}, Benjamin D. and {Ting}, Yuan-Sen and {Caldwell}, Nelson and {Zaritsky}, Dennis and {Cargile}, Phillip A.},
        title = "{Evidence from the H3 Survey That the Stellar Halo Is Entirely Comprised of Substructure}",
      journal = {\apj},
         year = 2020,
        month = sep,
       volume = {901},
       number = {1},
          eid = {48},
        pages = {48},
          doi = {10.3847/1538-4357/abaef4},
archivePrefix = {arXiv},
       eprint = {2006.08625},
 primaryClass = {astro-ph.GA},
       adsurl = {https://ui.adsabs.harvard.edu/abs/2020ApJ...901...48N}
}

@ARTICLE{Gibbons17,
       author = {{Gibbons}, S.~L.~J. and {Belokurov}, V. and {Evans}, N.~W.},
        title = "{A tail of two populations: chemo-dynamics of the Sagittarius stream and implications for its original mass}",
      journal = {\mnras},
         year = 2017,
        month = jan,
       volume = {464},
       number = {1},
        pages = {794-809},
          doi = {10.1093/mnras/stw2328},
archivePrefix = {arXiv},
       eprint = {1607.00803},
 primaryClass = {astro-ph.GA},
       adsurl = {https://ui.adsabs.harvard.edu/abs/2017MNRAS.464..794G}
}

@ARTICLE{Johnson20,
       author = {{Johnson}, Benjamin D. and {Conroy}, Charlie and {Naidu}, Rohan P. and {Bonaca}, Ana and {Zaritsky}, Dennis and {Ting}, Yuan-Sen and {Cargile}, Phillip A. and {Han}, Jiwon Jesse and {Speagle}, Joshua S.},
        title = "{A Diffuse Metal-poor Component of the Sagittarius Stream Revealed by the H3 Survey}",
      journal = {\apj},
         year = 2020,
        month = sep,
       volume = {900},
       number = {2},
          eid = {103},
        pages = {103},
          doi = {10.3847/1538-4357/abab08},
archivePrefix = {arXiv},
       eprint = {2007.14408},
 primaryClass = {astro-ph.GA},
       adsurl = {https://ui.adsabs.harvard.edu/abs/2020ApJ...900..103J}
}

@ARTICLE{Limberg23,
       author = {{Limberg}, Guilherme and {Queiroz}, Anna B.~A. and {Perottoni}, H{\'e}lio D. and {Rossi}, Silvia and {Amarante}, Jo{\~a}o A.~S. and {Santucci}, Rafael M. and {Chiappini}, Cristina and {P{\'e}rez-Villegas}, Angeles and {Lee}, Young Sun},
        title = "{Phase-space Properties and Chemistry of the Sagittarius Stellar Stream Down to the Extremely Metal-poor ([Fe/H] {\ensuremath{\lesssim}} -3) Regime}",
      journal = {\apj},
         year = 2023,
        month = apr,
       volume = {946},
       number = {2},
          eid = {66},
        pages = {66},
          doi = {10.3847/1538-4357/acb694},
archivePrefix = {arXiv},
       eprint = {2212.08249},
 primaryClass = {astro-ph.GA},
       adsurl = {https://ui.adsabs.harvard.edu/abs/2023ApJ...946...66L}
}

@ARTICLE{Bonaca25,
       author = {{Bonaca}, Ana and {Price-Whelan}, Adrian M.},
        title = "{Stellar streams in the Gaia era}",
      journal = {\nar},
         year = 2025,
        month = jun,
       volume = {100},
          eid = {101713},
        pages = {101713},
          doi = {10.1016/j.newar.2024.101713},
archivePrefix = {arXiv},
       eprint = {2405.19410},
 primaryClass = {astro-ph.GA},
       adsurl = {https://ui.adsabs.harvard.edu/abs/2025NewAR.10001713B}
}

@ARTICLE{Alard96,
       author = {{Alard}, C.},
        title = "{Evidence for the Sagittarius Dwarf Galaxy at Low Galactic Latitudes}",
      journal = {\apjl},
         year = 1996,
        month = feb,
       volume = {458},
        pages = {L17},
          doi = {10.1086/309917},
       adsurl = {https://ui.adsabs.harvard.edu/abs/1996ApJ...458L..17A}
}

@ARTICLE{Mateo96,
       author = {{Mateo}, Mario and {Mirabal}, Nestor and {Udalski}, A. and {Szymanski}, M. and {Kaluzny}, J. and {Kubiak}, M. and {Krzeminski}, W. and {Stanek}, K.~Z.},
        title = "{Discovery of a Tidal Extension of the Sagittarius Dwarf Spheroidal Galaxy}",
      journal = {\apjl},
         year = 1996,
        month = feb,
       volume = {458},
        pages = {L13},
          doi = {10.1086/309919},
       adsurl = {https://ui.adsabs.harvard.edu/abs/1996ApJ...458L..13M}
}

@ARTICLE{Mateo98,
       author = {{Mateo}, Mario and {Olszewski}, Edward W. and {Morrison}, Heather L.},
        title = "{Tracing the Outer Structure of the Sagittarius Dwarf Galaxy:Detections at Angular Distances between 10{\textdegree} and 34{\textdegree}}",
      journal = {\apjl},
         year = 1998,
        month = nov,
       volume = {508},
       number = {1},
        pages = {L55-L59},
          doi = {10.1086/311720},
archivePrefix = {arXiv},
       eprint = {astro-ph/9810015},
 primaryClass = {astro-ph},
       adsurl = {https://ui.adsabs.harvard.edu/abs/1998ApJ...508L..55M}
}

@ARTICLE{Ibata01,
       author = {{Ibata}, Rodrigo and {Irwin}, Michael and {Lewis}, Geraint F. and {Stolte}, Andrea},
        title = "{Galactic Halo Substructure in the Sloan Digital Sky Survey: The Ancient Tidal Stream from the Sagittarius Dwarf Galaxy}",
      journal = {\apjl},
         year = 2001,
        month = feb,
       volume = {547},
       number = {2},
        pages = {L133-L136},
          doi = {10.1086/318894},
archivePrefix = {arXiv},
       eprint = {astro-ph/0004255},
 primaryClass = {astro-ph},
       adsurl = {https://ui.adsabs.harvard.edu/abs/2001ApJ...547L.133I}
}

@ARTICLE{Newberg03,
       author = {{Newberg}, Heidi Jo and {Yanny}, Brian and {Grebel}, Eva K. and {Hennessy}, Greg and {Ivezi{\'c}}, {\v{Z}}eljko and {Martinez-Delgado}, David and {Odenkirchen}, Michael and {Rix}, Hans-Walter and {Brinkmann}, Jon and {Lamb}, Don Q. and {Schneider}, Donald P. and {York}, Donald G.},
        title = "{Sagittarius Tidal Debris 90 Kiloparsecs from the Galactic Center}",
      journal = {\apjl},
         year = 2003,
        month = oct,
       volume = {596},
       number = {2},
        pages = {L191-L194},
          doi = {10.1086/379316},
archivePrefix = {arXiv},
       eprint = {astro-ph/0309162},
 primaryClass = {astro-ph},
       adsurl = {https://ui.adsabs.harvard.edu/abs/2003ApJ...596L.191N}
}

@ARTICLE{Belokurov06,
       author = {{Belokurov}, V. and {Zucker}, D.~B. and {Evans}, N.~W. and {Gilmore}, G. and {Vidrih}, S. and {Bramich}, D.~M. and {Newberg}, H.~J. and {Wyse}, R.~F.~G. and {Irwin}, M.~J. and {Fellhauer}, M. and {Hewett}, P.~C. and {Walton}, N.~A. and {Wilkinson}, M.~I. and {Cole}, N. and {Yanny}, B. and {Rockosi}, C.~M. and {Beers}, T.~C. and {Bell}, E.~F. and {Brinkmann}, J. and {Ivezi{\'c}}, {\v{Z}}. and {Lupton}, R.},
        title = "{The Field of Streams: Sagittarius and Its Siblings}",
      journal = {\apjl},
         year = 2006,
        month = may,
       volume = {642},
       number = {2},
        pages = {L137-L140},
          doi = {10.1086/504797},
archivePrefix = {arXiv},
       eprint = {astro-ph/0605025},
 primaryClass = {astro-ph},
       adsurl = {https://ui.adsabs.harvard.edu/abs/2006ApJ...642L.137B}
}

@ARTICLE{Belokurov14,
       author = {{Belokurov}, V. and {Koposov}, S.~E. and {Evans}, N.~W. and {Pe{\~n}arrubia}, J. and {Irwin}, M.~J. and {Smith}, M.~C. and {Lewis}, G.~F. and {Gieles}, M. and {Wilkinson}, M.~I. and {Gilmore}, G. and {Olszewski}, E.~W. and {Niederste-Ostholt}, M.},
        title = "{Precession of the Sagittarius stream}",
      journal = {\mnras},
         year = 2014,
        month = jan,
       volume = {437},
       number = {1},
        pages = {116-131},
          doi = {10.1093/mnras/stt1862},
archivePrefix = {arXiv},
       eprint = {1301.7069},
 primaryClass = {astro-ph.GA},
       adsurl = {https://ui.adsabs.harvard.edu/abs/2014MNRAS.437..116B}
}

@ARTICLE{Hernitschek17,
       author = {{Hernitschek}, Nina and {Sesar}, Branimir and {Rix}, Hans-Walter and {Belokurov}, Vasily and {Martinez-Delgado}, David and {Martin}, Nicolas F. and {Kaiser}, Nick and {Hodapp}, Klaus and {Chambers}, Kenneth C. and {Wainscoat}, Richard and {Magnier}, Eugene and {Kudritzki}, Rolf-Peter and {Metcalfe}, Nigel and {Draper}, Peter W.},
        title = "{The Geometry of the Sagittarius Stream from Pan-STARRS1 3{\ensuremath{\pi}} RR Lyrae}",
      journal = {\apj},
         year = 2017,
        month = nov,
       volume = {850},
       number = {1},
          eid = {96},
        pages = {96},
          doi = {10.3847/1538-4357/aa960c},
archivePrefix = {arXiv},
       eprint = {1710.09436},
 primaryClass = {astro-ph.GA},
       adsurl = {https://ui.adsabs.harvard.edu/abs/2017ApJ...850...96H}
}

@ARTICLE{Helmi01,
       author = {{Helmi}, Amina and {White}, Simon D.~M.},
        title = "{Simple dynamical models of the Sagittarius dwarf galaxy}",
      journal = {\mnras},
         year = 2001,
        month = may,
       volume = {323},
       number = {3},
        pages = {529-536},
          doi = {10.1046/j.1365-8711.2001.04238.x},
archivePrefix = {arXiv},
       eprint = {astro-ph/0002482},
 primaryClass = {astro-ph},
       adsurl = {https://ui.adsabs.harvard.edu/abs/2001MNRAS.323..529H}
}

@ARTICLE{Helmi04,
       author = {{Helmi}, Amina},
        title = "{Velocity Trends in the Debris of Sagittarius and the Shape of the Dark Matter Halo of Our Galaxy}",
      journal = {\apjl},
         year = 2004,
        month = aug,
       volume = {610},
       number = {2},
        pages = {L97-L100},
          doi = {10.1086/423340},
archivePrefix = {arXiv},
       eprint = {astro-ph/0406396},
 primaryClass = {astro-ph},
       adsurl = {https://ui.adsabs.harvard.edu/abs/2004ApJ...610L..97H}
}

@ARTICLE{Johnston05,
       author = {{Johnston}, Kathryn V. and {Law}, David R. and {Majewski}, Steven R.},
        title = "{A Two Micron All Sky Survey View of the Sagittarius Dwarf Galaxy. III. Constraints on the Flattening of the Galactic Halo}",
      journal = {\apj},
         year = 2005,
        month = feb,
       volume = {619},
       number = {2},
        pages = {800-806},
          doi = {10.1086/426777},
archivePrefix = {arXiv},
       eprint = {astro-ph/0407565},
 primaryClass = {astro-ph},
       adsurl = {https://ui.adsabs.harvard.edu/abs/2005ApJ...619..800J}
}

@ARTICLE{Law05,
       author = {{Law}, David R. and {Johnston}, Kathryn V. and {Majewski}, Steven R.},
        title = "{A Two Micron All-Sky Survey View of the Sagittarius Dwarf Galaxy. IV. Modeling the Sagittarius Tidal Tails}",
      journal = {\apj},
         year = 2005,
        month = feb,
       volume = {619},
       number = {2},
        pages = {807-823},
          doi = {10.1086/426779},
archivePrefix = {arXiv},
       eprint = {astro-ph/0407566},
 primaryClass = {astro-ph},
       adsurl = {https://ui.adsabs.harvard.edu/abs/2005ApJ...619..807L}
}

@ARTICLE{Fellhauer06,
       author = {{Fellhauer}, M. and {Belokurov}, V. and {Evans}, N.~W. and {Wilkinson}, M.~I. and {Zucker}, D.~B. and {Gilmore}, G. and {Irwin}, M.~J. and {Bramich}, D.~M. and {Vidrih}, S. and {Wyse}, R.~F.~G. and {Beers}, T.~C. and {Brinkmann}, J.},
        title = "{The Origin of the Bifurcation in the Sagittarius Stream}",
      journal = {\apj},
         year = 2006,
        month = nov,
       volume = {651},
       number = {1},
        pages = {167-173},
          doi = {10.1086/507128},
archivePrefix = {arXiv},
       eprint = {astro-ph/0605026},
 primaryClass = {astro-ph},
       adsurl = {https://ui.adsabs.harvard.edu/abs/2006ApJ...651..167F}
}

@ARTICLE{Penarrubia10,
       author = {{Pe{\~n}arrubia}, Jorge and {Belokurov}, Vasily and {Evans}, N.~W. and {Mart{\'\i}nez-Delgado}, David and {Gilmore}, Gerard and {Irwin}, Mike and {Niederste-Ostholt}, Martin and {Zucker}, Daniel B.},
        title = "{Was the progenitor of the Sagittarius stream a disc galaxy?}",
      journal = {\mnras},
         year = 2010,
        month = oct,
       volume = {408},
       number = {1},
        pages = {L26-L30},
          doi = {10.1111/j.1745-3933.2010.00921.x},
archivePrefix = {arXiv},
       eprint = {1007.1485},
 primaryClass = {astro-ph.GA},
       adsurl = {https://ui.adsabs.harvard.edu/abs/2010MNRAS.408L..26P}
}

@ARTICLE{Law10,
       author = {{Law}, David R. and {Majewski}, Steven R.},
        title = "{The Sagittarius Dwarf Galaxy: A Model for Evolution in a Triaxial Milky Way Halo}",
      journal = {\apj},
         year = 2010,
        month = may,
       volume = {714},
       number = {1},
        pages = {229-254},
          doi = {10.1088/0004-637X/714/1/229},
archivePrefix = {arXiv},
       eprint = {1003.1132},
 primaryClass = {astro-ph.GA},
       adsurl = {https://ui.adsabs.harvard.edu/abs/2010ApJ...714..229L}
}

@ARTICLE{Gaia16a,
       author = {{Gaia Collaboration} and {Prusti}, T. and {de Bruijne}, J.~H.~J. and {Brown}, A.~G.~A. and {Vallenari}, A. and {Babusiaux}, C. and {Bailer-Jones}, C.~A.~L. and {Bastian}, U. and {Biermann}, M. and {Evans}, D.~W. and {Eyer}, L. and {Jansen}, F. and {Jordi}, C. and {Klioner}, S.~A. and {Lammers}, U. and {Lindegren}, L. and {Luri}, X. and {Mignard}, F. and {Milligan}, D.~J. and {Panem}, C. and {Poinsignon}, V. and {Pourbaix}, D. and {Randich}, S. and {Sarri}, G. and {Sartoretti}, P. and {Siddiqui}, H.~I. and {Soubiran}, C. and {Valette}, V. and {van Leeuwen}, F. and {Walton}, N.~A. and {Aerts}, C. and {Arenou}, F. and {Cropper}, M. and {Drimmel}, R. and {H{\o}g}, E. and {Katz}, D. and {Lattanzi}, M.~G. and {O'Mullane}, W. and {Grebel}, E.~K. and {Holland}, A.~D. and {Huc}, C. and {Passot}, X. and {Bramante}, L. and {Cacciari}, C. and {Casta{\~n}eda}, J. and {Chaoul}, L. and {Cheek}, N. and {De Angeli}, F. and {Fabricius}, C. and {Guerra}, R. and {Hern{\'a}ndez}, J. and {Jean-Antoine-Piccolo}, A. and {Masana}, E. and {Messineo}, R. and {Mowlavi}, N. and {Nienartowicz}, K. and {Ord{\'o}{\~n}ez-Blanco}, D. and {Panuzzo}, P. and {Portell}, J. and {Richards}, P.~J. and {Riello}, M. and {Seabroke}, G.~M. and {Tanga}, P. and {Th{\'e}venin}, F. and {Torra}, J. and {Els}, S.~G. and {Gracia-Abril}, G. and {Comoretto}, G. and {Garcia-Reinaldos}, M. and {Lock}, T. and {Mercier}, E. and {Altmann}, M. and {Andrae}, R. and {Astraatmadja}, T.~L. and {Bellas-Velidis}, I. and {Benson}, K. and {Berthier}, J. and {Blomme}, R. and {Busso}, G. and {Carry}, B. and {Cellino}, A. and {Clementini}, G. and {Cowell}, S. and {Creevey}, O. and {Cuypers}, J. and {Davidson}, M. and {De Ridder}, J. and {de Torres}, A. and {Delchambre}, L. and {Dell'Oro}, A. and {Ducourant}, C. and {Fr{\'e}mat}, Y. and {Garc{\'\i}a-Torres}, M. and {Gosset}, E. and {Halbwachs}, J.-L. and {Hambly}, N.~C. and {Harrison}, D.~L. and {Hauser}, M. and {Hestroffer}, D. and {Hodgkin}, S.~T. and {Huckle}, H.~E. and {Hutton}, A. and {Jasniewicz}, G. and {Jordan}, S. and {Kontizas}, M. and {Korn}, A.~J. and {Lanzafame}, A.~C. and {Manteiga}, M. and {Moitinho}, A. and {Muinonen}, K. and {Osinde}, J. and {Pancino}, E. and {Pauwels}, T. and {Petit}, J.-M. and {Recio-Blanco}, A. and {Robin}, A.~C. and {Sarro}, L.~M. and {Siopis}, C. and {Smith}, M. and {Smith}, K.~W. and {Sozzetti}, A. and {Thuillot}, W. and {van Reeven}, W. and {Viala}, Y. and {Abbas}, U. and {Abreu Aramburu}, A. and {Accart}, S. and {Aguado}, J.~J. and {Allan}, P.~M. and {Allasia}, W. and {Altavilla}, G. and {{\'A}lvarez}, M.~A. and {Alves}, J. and {Anderson}, R.~I. and {Andrei}, A.~H. and {Anglada Varela}, E. and {Antiche}, E. and {Antoja}, T. and {Ant{\'o}n}, S. and {Arcay}, B. and {Atzei}, A. and {Ayache}, L. and {Bach}, N. and {Baker}, S.~G. and {Balaguer-N{\'u}{\~n}ez}, L. and {Barache}, C. and {Barata}, C. and {Barbier}, A. and {Barblan}, F. and {Baroni}, M. and {Barrado y Navascu{\'e}s}, D. and {Barros}, M. and {Barstow}, M.~A. and {Becciani}, U. and {Bellazzini}, M. and {Bellei}, G. and {Bello Garc{\'\i}a}, A. and {Belokurov}, V. and {Bendjoya}, P. and {Berihuete}, A. and {Bianchi}, L. and {Bienaym{\'e}}, O. and {Billebaud}, F. and {Blagorodnova}, N. and {Blanco-Cuaresma}, S. and {Boch}, T. and {Bombrun}, A. and {Borrachero}, R. and {Bouquillon}, S. and {Bourda}, G. and {Bouy}, H. and {Bragaglia}, A. and {Breddels}, M.~A. and {Brouillet}, N. and {Br{\"u}semeister}, T. and {Bucciarelli}, B. and {Budnik}, F. and {Burgess}, P. and {Burgon}, R. and {Burlacu}, A. and {Busonero}, D. and {Buzzi}, R. and {Caffau}, E. and {Cambras}, J. and {Campbell}, H. and {Cancelliere}, R. and {Cantat-Gaudin}, T. and {Carlucci}, T. and {Carrasco}, J.~M. and {Castellani}, M. and {Charlot}, P. and {Charnas}, J. and {Charvet}, P. and {Chassat}, F. and {Chiavassa}, A. and {Clotet}, M. and {Cocozza}, G. and {Collins}, R.~S. and {Collins}, P. and {Costigan}, G.},
        title = "{The Gaia mission}",
      journal = {\aap},
         year = 2016,
        month = nov,
       volume = {595},
          eid = {A1},
        pages = {A1},
          doi = {10.1051/0004-6361/201629272},
archivePrefix = {arXiv},
       eprint = {1609.04153},
 primaryClass = {astro-ph.IM},
       adsurl = {https://ui.adsabs.harvard.edu/abs/2016A&A...595A...1G}
}

@ARTICLE{Gaia16b,
       author = {{Gaia Collaboration} and {Brown}, A.~G.~A. and {Vallenari}, A. and {Prusti}, T. and {de Bruijne}, J.~H.~J. and {Mignard}, F. and {Drimmel}, R. and {Babusiaux}, C. and {Bailer-Jones}, C.~A.~L. and {Bastian}, U. and {Biermann}, M. and {Evans}, D.~W. and {Eyer}, L. and {Jansen}, F. and {Jordi}, C. and {Katz}, D. and {Klioner}, S.~A. and {Lammers}, U. and {Lindegren}, L. and {Luri}, X. and {O'Mullane}, W. and {Panem}, C. and {Pourbaix}, D. and {Randich}, S. and {Sartoretti}, P. and {Siddiqui}, H.~I. and {Soubiran}, C. and {Valette}, V. and {van Leeuwen}, F. and {Walton}, N.~A. and {Aerts}, C. and {Arenou}, F. and {Cropper}, M. and {H{\o}g}, E. and {Lattanzi}, M.~G. and {Grebel}, E.~K. and {Holland}, A.~D. and {Huc}, C. and {Passot}, X. and {Perryman}, M. and {Bramante}, L. and {Cacciari}, C. and {Casta{\~n}eda}, J. and {Chaoul}, L. and {Cheek}, N. and {De Angeli}, F. and {Fabricius}, C. and {Guerra}, R. and {Hern{\'a}ndez}, J. and {Jean-Antoine-Piccolo}, A. and {Masana}, E. and {Messineo}, R. and {Mowlavi}, N. and {Nienartowicz}, K. and {Ord{\'o}{\~n}ez-Blanco}, D. and {Panuzzo}, P. and {Portell}, J. and {Richards}, P.~J. and {Riello}, M. and {Seabroke}, G.~M. and {Tanga}, P. and {Th{\'e}venin}, F. and {Torra}, J. and {Els}, S.~G. and {Gracia-Abril}, G. and {Comoretto}, G. and {Garcia-Reinaldos}, M. and {Lock}, T. and {Mercier}, E. and {Altmann}, M. and {Andrae}, R. and {Astraatmadja}, T.~L. and {Bellas-Velidis}, I. and {Benson}, K. and {Berthier}, J. and {Blomme}, R. and {Busso}, G. and {Carry}, B. and {Cellino}, A. and {Clementini}, G. and {Cowell}, S. and {Creevey}, O. and {Cuypers}, J. and {Davidson}, M. and {De Ridder}, J. and {de Torres}, A. and {Delchambre}, L. and {Dell'Oro}, A. and {Ducourant}, C. and {Fr{\'e}mat}, Y. and {Garc{\'\i}a-Torres}, M. and {Gosset}, E. and {Halbwachs}, J.-L. and {Hambly}, N.~C. and {Harrison}, D.~L. and {Hauser}, M. and {Hestroffer}, D. and {Hodgkin}, S.~T. and {Huckle}, H.~E. and {Hutton}, A. and {Jasniewicz}, G. and {Jordan}, S. and {Kontizas}, M. and {Korn}, A.~J. and {Lanzafame}, A.~C. and {Manteiga}, M. and {Moitinho}, A. and {Muinonen}, K. and {Osinde}, J. and {Pancino}, E. and {Pauwels}, T. and {Petit}, J.-M. and {Recio-Blanco}, A. and {Robin}, A.~C. and {Sarro}, L.~M. and {Siopis}, C. and {Smith}, M. and {Smith}, K.~W. and {Sozzetti}, A. and {Thuillot}, W. and {van Reeven}, W. and {Viala}, Y. and {Abbas}, U. and {Abreu Aramburu}, A. and {Accart}, S. and {Aguado}, J.~J. and {Allan}, P.~M. and {Allasia}, W. and {Altavilla}, G. and {{\'A}lvarez}, M.~A. and {Alves}, J. and {Anderson}, R.~I. and {Andrei}, A.~H. and {Anglada Varela}, E. and {Antiche}, E. and {Antoja}, T. and {Ant{\'o}n}, S. and {Arcay}, B. and {Bach}, N. and {Baker}, S.~G. and {Balaguer-N{\'u}{\~n}ez}, L. and {Barache}, C. and {Barata}, C. and {Barbier}, A. and {Barblan}, F. and {Barrado y Navascu{\'e}s}, D. and {Barros}, M. and {Barstow}, M.~A. and {Becciani}, U. and {Bellazzini}, M. and {Bello Garc{\'\i}a}, A. and {Belokurov}, V. and {Bendjoya}, P. and {Berihuete}, A. and {Bianchi}, L. and {Bienaym{\'e}}, O. and {Billebaud}, F. and {Blagorodnova}, N. and {Blanco-Cuaresma}, S. and {Boch}, T. and {Bombrun}, A. and {Borrachero}, R. and {Bouquillon}, S. and {Bourda}, G. and {Bouy}, H. and {Bragaglia}, A. and {Breddels}, M.~A. and {Brouillet}, N. and {Br{\"u}semeister}, T. and {Bucciarelli}, B. and {Burgess}, P. and {Burgon}, R. and {Burlacu}, A. and {Busonero}, D. and {Buzzi}, R. and {Caffau}, E. and {Cambras}, J. and {Campbell}, H. and {Cancelliere}, R. and {Cantat-Gaudin}, T. and {Carlucci}, T. and {Carrasco}, J.~M. and {Castellani}, M. and {Charlot}, P. and {Charnas}, J. and {Chiavassa}, A. and {Clotet}, M. and {Cocozza}, G. and {Collins}, R.~S. and {Costigan}, G. and {Crifo}, F. and {Cross}, N.~J.~G. and {Crosta}, M. and {Crowley}, C. and {Dafonte}, C. and {Damerdji}, Y. and {Dapergolas}, A. and {David}, P. and {David}, M. and {De Cat}, P.},
        title = "{Gaia Data Release 1. Summary of the astrometric, photometric, and survey properties}",
      journal = {\aap},
         year = 2016,
        month = nov,
       volume = {595},
          eid = {A2},
        pages = {A2},
          doi = {10.1051/0004-6361/201629512},
archivePrefix = {arXiv},
       eprint = {1609.04172},
 primaryClass = {astro-ph.IM},
       adsurl = {https://ui.adsabs.harvard.edu/abs/2016A&A...595A...2G}
}

@ARTICLE{Gaia18,
       author = {{Gaia Collaboration} and {Brown}, A.~G.~A. and {Vallenari}, A. and {Prusti}, T. and {de Bruijne}, J.~H.~J. and {Babusiaux}, C. and {Bailer-Jones}, C.~A.~L. and {Biermann}, M. and {Evans}, D.~W. and {Eyer}, L. and {Jansen}, F. and {Jordi}, C. and {Klioner}, S.~A. and {Lammers}, U. and {Lindegren}, L. and {Luri}, X. and {Mignard}, F. and {Panem}, C. and {Pourbaix}, D. and {Randich}, S. and {Sartoretti}, P. and {Siddiqui}, H.~I. and {Soubiran}, C. and {van Leeuwen}, F. and {Walton}, N.~A. and {Arenou}, F. and {Bastian}, U. and {Cropper}, M. and {Drimmel}, R. and {Katz}, D. and {Lattanzi}, M.~G. and {Bakker}, J. and {Cacciari}, C. and {Casta{\~n}eda}, J. and {Chaoul}, L. and {Cheek}, N. and {De Angeli}, F. and {Fabricius}, C. and {Guerra}, R. and {Holl}, B. and {Masana}, E. and {Messineo}, R. and {Mowlavi}, N. and {Nienartowicz}, K. and {Panuzzo}, P. and {Portell}, J. and {Riello}, M. and {Seabroke}, G.~M. and {Tanga}, P. and {Th{\'e}venin}, F. and {Gracia-Abril}, G. and {Comoretto}, G. and {Garcia-Reinaldos}, M. and {Teyssier}, D. and {Altmann}, M. and {Andrae}, R. and {Audard}, M. and {Bellas-Velidis}, I. and {Benson}, K. and {Berthier}, J. and {Blomme}, R. and {Burgess}, P. and {Busso}, G. and {Carry}, B. and {Cellino}, A. and {Clementini}, G. and {Clotet}, M. and {Creevey}, O. and {Davidson}, M. and {De Ridder}, J. and {Delchambre}, L. and {Dell'Oro}, A. and {Ducourant}, C. and {Fern{\'a}ndez-Hern{\'a}ndez}, J. and {Fouesneau}, M. and {Fr{\'e}mat}, Y. and {Galluccio}, L. and {Garc{\'\i}a-Torres}, M. and {Gonz{\'a}lez-N{\'u}{\~n}ez}, J. and {Gonz{\'a}lez-Vidal}, J.~J. and {Gosset}, E. and {Guy}, L.~P. and {Halbwachs}, J.-L. and {Hambly}, N.~C. and {Harrison}, D.~L. and {Hern{\'a}ndez}, J. and {Hestroffer}, D. and {Hodgkin}, S.~T. and {Hutton}, A. and {Jasniewicz}, G. and {Jean-Antoine-Piccolo}, A. and {Jordan}, S. and {Korn}, A.~J. and {Krone-Martins}, A. and {Lanzafame}, A.~C. and {Lebzelter}, T. and {L{\"o}ffler}, W. and {Manteiga}, M. and {Marrese}, P.~M. and {Mart{\'\i}n-Fleitas}, J.~M. and {Moitinho}, A. and {Mora}, A. and {Muinonen}, K. and {Osinde}, J. and {Pancino}, E. and {Pauwels}, T. and {Petit}, J.-M. and {Recio-Blanco}, A. and {Richards}, P.~J. and {Rimoldini}, L. and {Robin}, A.~C. and {Sarro}, L.~M. and {Siopis}, C. and {Smith}, M. and {Sozzetti}, A. and {S{\"u}veges}, M. and {Torra}, J. and {van Reeven}, W. and {Abbas}, U. and {Abreu Aramburu}, A. and {Accart}, S. and {Aerts}, C. and {Altavilla}, G. and {{\'A}lvarez}, M.~A. and {Alvarez}, R. and {Alves}, J. and {Anderson}, R.~I. and {Andrei}, A.~H. and {Anglada Varela}, E. and {Antiche}, E. and {Antoja}, T. and {Arcay}, B. and {Astraatmadja}, T.~L. and {Bach}, N. and {Baker}, S.~G. and {Balaguer-N{\'u}{\~n}ez}, L. and {Balm}, P. and {Barache}, C. and {Barata}, C. and {Barbato}, D. and {Barblan}, F. and {Barklem}, P.~S. and {Barrado}, D. and {Barros}, M. and {Barstow}, M.~A. and {Bartholom{\'e} Mu{\~n}oz}, S. and {Bassilana}, J.-L. and {Becciani}, U. and {Bellazzini}, M. and {Berihuete}, A. and {Bertone}, S. and {Bianchi}, L. and {Bienaym{\'e}}, O. and {Blanco-Cuaresma}, S. and {Boch}, T. and {Boeche}, C. and {Bombrun}, A. and {Borrachero}, R. and {Bossini}, D. and {Bouquillon}, S. and {Bourda}, G. and {Bragaglia}, A. and {Bramante}, L. and {Breddels}, M.~A. and {Bressan}, A. and {Brouillet}, N. and {Br{\"u}semeister}, T. and {Brugaletta}, E. and {Bucciarelli}, B. and {Burlacu}, A. and {Busonero}, D. and {Butkevich}, A.~G. and {Buzzi}, R. and {Caffau}, E. and {Cancelliere}, R. and {Cannizzaro}, G. and {Cantat-Gaudin}, T. and {Carballo}, R. and {Carlucci}, T. and {Carrasco}, J.~M. and {Casamiquela}, L. and {Castellani}, M. and {Castro-Ginard}, A. and {Charlot}, P. and {Chemin}, L. and {Chiavassa}, A. and {Cocozza}, G. and {Costigan}, G. and {Cowell}, S. and {Crifo}, F. and {Crosta}, M. and {Crowley}, C. and {Cuypers}, J. and {Dafonte}, C. and {Damerdji}, Y. and {Dapergolas}, A. and {David}, P. and {David}, M. and {de Laverny}, P. and {De Luise}, F.},
        title = "{Gaia Data Release 2. Summary of the contents and survey properties}",
      journal = {\aap},
         year = 2018,
        month = aug,
       volume = {616},
          eid = {A1},
        pages = {A1},
          doi = {10.1051/0004-6361/201833051},
archivePrefix = {arXiv},
       eprint = {1804.09365},
 primaryClass = {astro-ph.GA},
       adsurl = {https://ui.adsabs.harvard.edu/abs/2018A&A...616A...1G}
}

@ARTICLE{Gaia21,
       author = {{Gaia Collaboration} and {Brown}, A.~G.~A. and {Vallenari}, A. and {Prusti}, T. and {de Bruijne}, J.~H.~J. and {Babusiaux}, C. and {Biermann}, M. and {Creevey}, O.~L. and {Evans}, D.~W. and {Eyer}, L. and {Hutton}, A. and {Jansen}, F. and {Jordi}, C. and {Klioner}, S.~A. and {Lammers}, U. and {Lindegren}, L. and {Luri}, X. and {Mignard}, F. and {Panem}, C. and {Pourbaix}, D. and {Randich}, S. and {Sartoretti}, P. and {Soubiran}, C. and {Walton}, N.~A. and {Arenou}, F. and {Bailer-Jones}, C.~A.~L. and {Bastian}, U. and {Cropper}, M. and {Drimmel}, R. and {Katz}, D. and {Lattanzi}, M.~G. and {van Leeuwen}, F. and {Bakker}, J. and {Cacciari}, C. and {Casta{\~n}eda}, J. and {De Angeli}, F. and {Ducourant}, C. and {Fabricius}, C. and {Fouesneau}, M. and {Fr{\'e}mat}, Y. and {Guerra}, R. and {Guerrier}, A. and {Guiraud}, J. and {Jean-Antoine Piccolo}, A. and {Masana}, E. and {Messineo}, R. and {Mowlavi}, N. and {Nicolas}, C. and {Nienartowicz}, K. and {Pailler}, F. and {Panuzzo}, P. and {Riclet}, F. and {Roux}, W. and {Seabroke}, G.~M. and {Sordo}, R. and {Tanga}, P. and {Th{\'e}venin}, F. and {Gracia-Abril}, G. and {Portell}, J. and {Teyssier}, D. and {Altmann}, M. and {Andrae}, R. and {Bellas-Velidis}, I. and {Benson}, K. and {Berthier}, J. and {Blomme}, R. and {Brugaletta}, E. and {Burgess}, P.~W. and {Busso}, G. and {Carry}, B. and {Cellino}, A. and {Cheek}, N. and {Clementini}, G. and {Damerdji}, Y. and {Davidson}, M. and {Delchambre}, L. and {Dell'Oro}, A. and {Fern{\'a}ndez-Hern{\'a}ndez}, J. and {Galluccio}, L. and {Garc{\'\i}a-Lario}, P. and {Garcia-Reinaldos}, M. and {Gonz{\'a}lez-N{\'u}{\~n}ez}, J. and {Gosset}, E. and {Haigron}, R. and {Halbwachs}, J.-L. and {Hambly}, N.~C. and {Harrison}, D.~L. and {Hatzidimitriou}, D. and {Heiter}, U. and {Hern{\'a}ndez}, J. and {Hestroffer}, D. and {Hodgkin}, S.~T. and {Holl}, B. and {Jan{\ss}en}, K. and {Jevardat de Fombelle}, G. and {Jordan}, S. and {Krone-Martins}, A. and {Lanzafame}, A.~C. and {L{\"o}ffler}, W. and {Lorca}, A. and {Manteiga}, M. and {Marchal}, O. and {Marrese}, P.~M. and {Moitinho}, A. and {Mora}, A. and {Muinonen}, K. and {Osborne}, P. and {Pancino}, E. and {Pauwels}, T. and {Petit}, J.-M. and {Recio-Blanco}, A. and {Richards}, P.~J. and {Riello}, M. and {Rimoldini}, L. and {Robin}, A.~C. and {Roegiers}, T. and {Rybizki}, J. and {Sarro}, L.~M. and {Siopis}, C. and {Smith}, M. and {Sozzetti}, A. and {Ulla}, A. and {Utrilla}, E. and {van Leeuwen}, M. and {van Reeven}, W. and {Abbas}, U. and {Abreu Aramburu}, A. and {Accart}, S. and {Aerts}, C. and {Aguado}, J.~J. and {Ajaj}, M. and {Altavilla}, G. and {{\'A}lvarez}, M.~A. and {{\'A}lvarez Cid-Fuentes}, J. and {Alves}, J. and {Anderson}, R.~I. and {Anglada Varela}, E. and {Antoja}, T. and {Audard}, M. and {Baines}, D. and {Baker}, S.~G. and {Balaguer-N{\'u}{\~n}ez}, L. and {Balbinot}, E. and {Balog}, Z. and {Barache}, C. and {Barbato}, D. and {Barros}, M. and {Barstow}, M.~A. and {Bartolom{\'e}}, S. and {Bassilana}, J.-L. and {Bauchet}, N. and {Baudesson-Stella}, A. and {Becciani}, U. and {Bellazzini}, M. and {Bernet}, M. and {Bertone}, S. and {Bianchi}, L. and {Blanco-Cuaresma}, S. and {Boch}, T. and {Bombrun}, A. and {Bossini}, D. and {Bouquillon}, S. and {Bragaglia}, A. and {Bramante}, L. and {Breedt}, E. and {Bressan}, A. and {Brouillet}, N. and {Bucciarelli}, B. and {Burlacu}, A. and {Busonero}, D. and {Butkevich}, A.~G. and {Buzzi}, R. and {Caffau}, E. and {Cancelliere}, R. and {C{\'a}novas}, H. and {Cantat-Gaudin}, T. and {Carballo}, R. and {Carlucci}, T. and {Carnerero}, M.~I. and {Carrasco}, J.~M. and {Casamiquela}, L. and {Castellani}, M. and {Castro-Ginard}, A. and {Castro Sampol}, P. and {Chaoul}, L. and {Charlot}, P. and {Chemin}, L. and {Chiavassa}, A. and {Cioni}, M.-R.~L. and {Comoretto}, G. and {Cooper}, W.~J. and {Cornez}, T. and {Cowell}, S. and {Crifo}, F. and {Crosta}, M. and {Crowley}, C. and {Dafonte}, C. and {Dapergolas}, A. and {David}, M. and {David}, P.},
        title = "{Gaia Early Data Release 3. Summary of the contents and survey properties}",
      journal = {\aap},
         year = 2021,
        month = may,
       volume = {649},
          eid = {A1},
        pages = {A1},
          doi = {10.1051/0004-6361/202039657},
archivePrefix = {arXiv},
       eprint = {2012.01533},
 primaryClass = {astro-ph.GA},
       adsurl = {https://ui.adsabs.harvard.edu/abs/2021A&A...649A...1G}
}

@ARTICLE{Antoja20,
       author = {{Antoja}, T. and {Ramos}, P. and {Mateu}, C. and {Helmi}, A. and {Anders}, F. and {Jordi}, C. and {Carballo-Bello}, J.~A.},
        title = "{An all-sky proper-motion map of the Sagittarius stream using Gaia DR2}",
      journal = {\aap},
         year = 2020,
        month = mar,
       volume = {635},
          eid = {L3},
        pages = {L3},
          doi = {10.1051/0004-6361/201937145},
archivePrefix = {arXiv},
       eprint = {2001.10012},
 primaryClass = {astro-ph.GA},
       adsurl = {https://ui.adsabs.harvard.edu/abs/2020A&A...635L...3A}
}
\bibliographystyle{aasjournalv7}

\end{document}